\documentclass[11pt,a4paper]{article}

\usepackage{epsfig}
\usepackage{amssymb}
\usepackage{array}
\usepackage{amsmath}
\usepackage{graphicx}

\newcommand{\ncmd}{\newcommand}

\newtheorem{defi}{Definition}
\newtheorem{theo}{Theorem}
\newtheorem{prop}{Proposition}
\newtheorem{lem}{Lemma}
\newtheorem{cor}{Corollary}

\newtheorem{rem}{Remark}

\newcommand{\be}{\begin{equation}}
\newcommand{\ee}{\end{equation}}
\newcommand{\bea}{\begin{eqnarray}}
\newcommand{\eea}{\end{eqnarray}}
\newcommand{\nn}{\nonumber}
\ncmd{\btheo}{\begin{theo}$\!\!\!$. -- }
\ncmd{\etheo}{\end{theo}}
\ncmd{\bpro}{\begin{prop}$\!\!\!$. -- }
\ncmd{\epro}{\end{prop}}
\ncmd{\preuve}{{\sc Preuve --}\ }
\ncmd{\bdefi}{\begin{defi} $\!\!\!$. -- }
\ncmd{\edefi}{\end{defi}}
\ncmd{\bco}{\begin{cor}$\!\!\!$. -- }
\ncmd{\eco}{\end{cor}}
\ncmd{\ble}{\begin{lem}$\!\!\!$. -- }
\ncmd{\ele}{\end{lem}}
\ncmd{\bno}{\begin{notation}$\!\!\!\!\!$. -- }
\ncmd{\eno}{\end{notation}}
\ncmd{\bre}{\begin{rem}$\!\!\!$. --  \begin{em}}
\ncmd{\ere}{\end{em} \end{rem}}
\ncmd{\beq}{\begin{equation}}
\ncmd{\eeq}{\end{equation}}
\ncmd{\ben}{\begin{enumerate}}
\ncmd{\een}{\end{enumerate}}
\ncmd{\bit}{\begin{itemize}}
\ncmd{\eit}{\end{itemize}}
\ncmd{\refp}[1]{(\ref{#1})}

\ncmd{\Fi}{\mathbb{F}}
\ncmd{\Oc}{\mathbb{O}}
\ncmd{\Ha}{\mathbb{H}}
\ncmd{\R}{\mathbb{R}}
\ncmd{\C}{\mathbb{C}}
\ncmd{\Z}{\mathbb{Z}}
\ncmd{\N}{\mathbb{N}}
\ncmd{\Sph}{\mathbb{S}}
\ncmd{\T}{\mathbb{T}}
\ncmd{\D}{\mathbb{D}}
\ncmd{\Lp}{\mathfrak{p}}
\ncmd{\Lg}{\mathfrak{g}}
\ncmd{\La}{\mathfrak{a}}
\ncmd{\Lk}{\mathfrak{k}}
\ncmd{\Lm}{\mathfrak{m}}
\ncmd{\Lh}{\mathfrak{h}}
\ncmd{\tr}{\mbox{tr}}
\ncmd{\ad}{\mbox{ad}}
\ncmd{\Ad}{\mbox{Ad}}
\ncmd{\diff}{\mbox{Diff}(M)}
\ncmd{\End}{\mbox{End}}
\ncmd{\Exp}{\mbox{Exp}}
\ncmd{\HH}{\mbox{H}}
\ncmd{\V}{\mbox{V}}
\ncmd{\riem}{\mbox{Riem}(M)}
\ncmd{\Aut}{\mbox{Aut}}
\ncmd{\Id}{\mbox{\tiny Id}}
\ncmd{\I}{\mathcal{I}}
\ncmd{\Ker}{\mbox{Ker}}
\ncmd{\di}{\displaystyle}
\ncmd{\bs}{\backslash}
\ncmd{\ov}{\overline}
\ncmd{\no}{\noindent}
\ncmd{\ra}{\rightarrow}
\ncmd{\lra}{\longrightarrow}
\ncmd{\eps}{\epsilon}
\ncmd{\M}{\mathcal{M}}
\ncmd{\DD}{\mathcal{D}}
\ncmd{\super}{\mathcal{S}}
\numberwithin{equation}{section}

\ncmd{\scalar}[2]{\mbox{$\mathcal{h} #1,#2 \mathcal{i}$}}
\ncmd{\ichap}{\^{\i}}

\title{Classical Gauge Theory in Riem}
\author{Henrique de A. Gomes\footnote{University of Nottingham, School of Mathematical Sciences, gomes.ha@gmail.com, pmxhg3@nottingham.ac.uk
}}

\begin{document}
\maketitle

\begin{abstract}
In the  geometrodynamical setting of general relativity in Lagrangian form, the objects of study are the
  {\it Riemannian}  metrics (and their time derivatives) over a given 3-manifold $M$.  It is our aim in this paper to study some geometrical aspects of the space $\M:=$Riem$(M)$ of all metrics over $M$.  For instance, the Hamiltonian constraints by themselves do not generate a group, and thus its action on $\riem$ cannot be viewed in a geometrical gauge setting. It is possible to do so for the momentum constraints however. Furthermore, in view of the recent results representing GR as a dual theory, invariant under foliation preserving 3--diffeomorphisms and 3D conformal transformations, but not under refoliations,  we are justified in considering the gauge structure pertaining only to the groups $\DD$ of diffeomorphisms of $M$, and $\mathcal{C}$, of conformal diffeomorphisms on $M$.
  For these infinite-dimensional symmetry groups, $\M$ has a
  natural principal fiber bundle (PFB) structure,
  which renders the gravitational field amenable to the full range of gauge-theoretic treatment. The aim of the paper is to use the geometrical structure present in the configuration space of general relativity to build gauge connection forms. The interpretation of the gauge connection form for the 3-diffeomorphism group is that it yields parallel translation of coordinates. For the conformal group, it yields parallel translation of scale.
We focus on the concept of a gauge connection forms for these structures and construct explicit formulae for supermetric-induced
    gauge connections.
    To apply the formalism, we compute general properties for a specific connection
   bearing strong resemblance to the one naturally induced by the deWitt supermetric,
    showing it has desirable relationalist properties.
   \end{abstract}
\newpage
\tableofcontents

\section{Introduction}

Gauge theory, needless to say, has a long and rich history, and it is probably not an exaggeration to state it has by now permeated all areas of theoretical physics as an essential tool for existing frameworks and guide for future developments. It describes systems which possess some inherent symmetry in their parametrizations, and for classical fields over spacetime it has a well-developed geometrical understanding through the use of principal fiber bundles.

Geometrodynamics, as championed by Wheeler, is the study of gravitation through a primary focus on {\it space and
changes therein} rather than on space-time itself. Space-time is essentially `sliced-up' and described as an
evolution of the geometry of these spatial slices through time. It is fundamentally a dynamical view of GR,
 technically taking form as its constrained canonical, or  ADM formulation \cite{Arnowitt:1962hi}.

 Although widely regarded as a gauge theory (since all of its constraints are first class and thus interpreted as symmetry generating), there is no specific description of ADM as a gauge theory in the geometric, fiber bundle sense, making use of connection forms, sections and so forth. This is in part because a connection over configuration space seems to be far removed from reality. What would such a connection do? This is one of the questions we aim to answer in this part.

As is well known,  the unconstrained configuration space for General Relativity is defined as
$$\M:=\riem=~~~\mbox{the space of all 3-Riemannian metrics over }~~ M~$$
The Hamiltonian dynamics thus takes place on (a constraint submanifold of) $T^*\riem$. By a geometrical setting of gauge theory, mathematically we mean the existence of a principal fiber bundle and, most importantly, a connection form on it.
The first inkling of a connection form in $\riem $ arose in \cite{Giulini:1993ct,Giulini:2009np}, where the mention of horizontal and vertical components of metric velocities first appears. It is however our understanding that the concept was not fully explored, and one of the purposes of the present work, at least from the mathematical standpoint, is to investigate exactly what constitutes a connection over configuration space. That is,  what are the properties such a connection has to satisfy and how can we construct one both formally and explicitly. In doing so we would like to shed light on explicit infinite-dimensional geometrical gauge theories over configuration space, a point of view so far as we know original.
From the physical point of view, this may be connected with a richer history of relational ideas \cite{Barbour:shape_dynamics}.

A second reason why this approach has not been attempted before is because of the difficulties in interpreting the action of the Hamiltonian constraint as a group action, and other issues related to the infamous ``problem of time" \cite{Anderson:2010xm}. As in the first part of the thesis we have shown that it is possible to have a theory of gravity which no longer possesses the scalar (or Hamiltonian)  constraint, and thus no refoliation invariance. Unlike what is the case in ADM, the constraints of this dual theory then form subalgebras, reflecting the kind of group structure suitable for an exhaustive principal fiber bundle formulation. This points in a new direction for the development of  gauge theoretic tools for gravity and sets the stage for applying more standard methods for the quantization of gravity as a gauge theory.

 Motivated by the possibility of now describing the symmetry groups of general relativity in a full geometrical gauge-theoretic setting, we will attempt to make explicit the  gauge connections relating  to the action of these two groups; the group of three dimensional diffeomorphisms, which we denote by $\mathcal{D}$, and that of three-dimensional conformal transformations $\mathcal{C}$.  Both $\mathcal{C}$ and $\mathcal{D}$ groups have right actions on the natural configuration space $\M$.
 We will constrain our attention to the case of $M$ being compact and closed, which is of more interest to the relational approach for various reasons \cite{Barbour94}.

 \subsection{Principal fiber bundles and gauge theory.}

\begin{figure}\label{fig:PFB}
 \begin{center}
 \includegraphics[width=11cm]{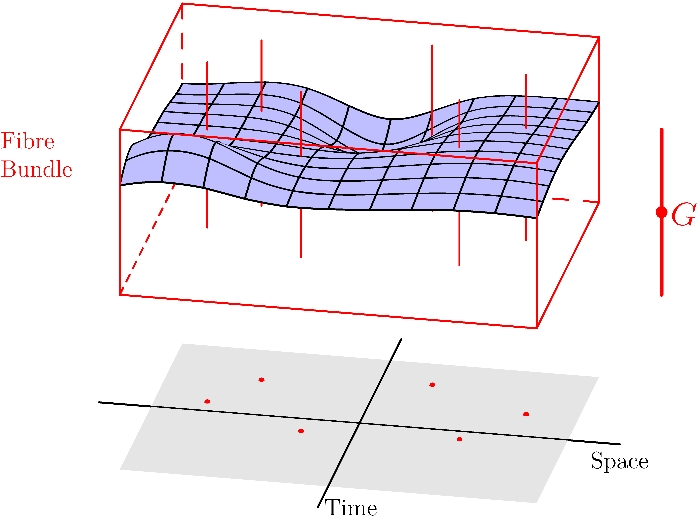}

\caption{A section in a principal fiber bundle over space-time.}

 \end{center}

\end{figure}

 Here we will briefly introduce the concept of a principal fiber-bundle, depicted in figure \ref{fig:PFB}. We will first present the formal definitions and

\begin{defi}\label{def:PFB}
A principal fiber bundle with smooth structural group $G$ is a smooth manifold $P$ on which $G$ acts $G\times P\rightarrow P$ and for which the action of $G$ is smooth and free. By a free action we mean that
$$G_p=\{h\in{G} ~|~ hp=p\}=\{\mbox{Id}\}$$
That is, the isotropy group of every point is the identity.
\end{defi}
One then constructs a projection
\be \mbox{pr}:P\rightarrow P/G=:B
\ee
where the base manifold $B$ is defined with the quotient topology with respect to the equivalence relation $p\simeq{q}\Leftrightarrow{p=h\cdot{q}}$, for some  $h\in G$. We call an orbit of $p\in P$ (or of $\mbox{pr}(p)=x\in B$) a \emph{fiber}, and also denote it by  $\mathcal{O}_x:=\mbox{pr}^{-1}(x)$.

For finite-dimensional manifolds, by the freedom of the group action, we can see that the orbits are isomorphic to the group $G$, but have no preferred identity element.\footnote{Such objects are in modern mathematical language called $G$-torsors. }\label{footn3} A smooth choice of identity element in each fiber coincides with the definition of a local section:
  \begin{defi}
Let $U$ be an open set in $B$. We define a \emph{local section} of $P$ over  $U$ as a submanifold
$\Sigma$ of $P$ such that for every $x\in U$,  $\Sigma$  is transversal to the orbits,~ $T_p\Sigma\oplus T_p\mathcal{O}_x=T_pP$,~ and $\Sigma$ intersects orbits over $U$ at a single point; i.e. for $p\in\Sigma$ then $\mathcal{O}_p\cap\Sigma=\{p\}$
  \end{defi}
 In rough terms, this means that a section $i)$ never has a component along the orbits and intersects every orbit (transversality) $ii)$ that it intersects each orbit once. These facts are enough to show that we can completely characterize an element of $P$ over $U$ by its location on the base and an element of the group which says where it is wrt the section. A choice of section is also called a \emph{choice of local gauge}. One can prove from this definition that in finite dimensions there always exists a \emph{slice} for our definition of a principal fiber bundle (definition \ref{def:PFB})  (see \cite{Palais}\footnote{Or see Theorem 19 in \cite{gomes-dissertation} for a detailed proof in the language used in section \ref{sec:slice}.}).  Indeed, it is using the same outlines of that proof that we are able to show that a section for the action of different groups on Riem exists (see section \ref{sec:slice}). A slice implies that $P$ has a \emph{local product structure} that we can patch  together to form an atlas of the manifold, and that all slices over the same open set are diffeomorphic. These \emph{sections} are then equivalent to the concept of {a gauge}, and transition maps from one gauge (or section) to the other can be shown to be functions $\Psi_{UU'}:U\cap U'\rightarrow G$.

 \subsection{Example: the bundle of bases.}

 The simplest and most telling example of a principal fiber bundle, is the one of all linear bases of $TM$, for a given manifold $M$. The group $\mbox{GL}(n)$ acts smoothly and has trivial isotropy, meaning it doesn't act trivially on any base. There is no preferred identity element (a preferred basis of each tangent space), and yet we can take every base to every other base by an action of $\mbox{GL}(n)$, making each fiber isomorphic to $\mbox{GL}(n)$. It is useful for us to already preview the concept of a \emph{connection} in this setting. Given a base $e$ over the point $x\in M$ and a vector $v\in T_xM$, a \emph{connection} will basically tell us which base corresponds to $e$ in that given direction, i.e., how to define parallel transport of the basis $e$ in each direction.
 
 \subsection{Notation}

Throughout the paper  semi-colon denotes covariant differentiation, and we
will, when it is convenient, use abstract index notation
(parentheses denote symmetrization of indices, and square brackets
anti-symmetrization). Also, again when it is convenient, we shall use $\nabla_a$
to denote the intrinsic Levi-Civita covariant derivative related to the
3-metric, and $D_a$ the one related to the 4-dimensional one.

The one parameter family of natural metrics on the tangent space to Riem (the
configuration space of all 3-metrics) is taken to be given by
\cite{Giulini:1993ct}:
 \be\label{supermetric} \mathcal{G}_\beta(u,v)_g=\int_M
G^{abcd}_\beta u_{ab}v_{cd} d\mu_g,\ee where, for tangent vectors $u,v\in
T_g(\mbox{Riem})$, the \emph{generalized DeWitt metric} is defined as
   \be\label{deWitt}
G_{\beta}^{abcd}:=g^{ac}g^{bd}-\beta g^{ab}g^{cd}
   \ee
{with inverse}
\be G^{\beta}_{abcd}:=g^{ac}g^{bd}-\lambda g^{ab}g^{cd},\ee where by inverse we
mean
$G_{\beta}^{abnm}G^{\beta}_{cdnm}=\delta^a_c\delta^b_d$. The relation between $\beta$ and $\lambda$ is that $\lambda=\frac{\beta}{3\beta-1} $. The usual
 DeWitt metric is $G_1$.
  We briefly note that the DeWitt metric is usually taken to be
  $(\sqrt{g}/2)(g^{ac}g^{bd}+g^{ad}g^{bc}-2g^{ab}g^{cd})$, but if we are only dealing with
   symmetric two-valence tensors, its action amounts to the one we have used,
apart from
   the $\sqrt{g}$ factor, which we input on the volume form.

\section{Riem as a principal fiber bundle}\label{sec:Riem}

This section includes only the basic tools that allows us to regard Riem as a principal fiber bundle (with with different structure groups). Mathematical subtleties arise as we are dealing with function spaces. These subtleties are however well under control, and for the more technical treatment and proofs we refer the reader to the appendix. 

\subsection{The 3-diffeomorphism group.}\label{sec:3-diffeo}

Let $E=S^2T^*:=TM^*\otimes_STM^*$ denote the symmetric product of the cotangent bundle, and $\Gamma^\infty(S^2T^*)$ the
space of smooth sections over this bundle\footnote{It is a  Frech\'et space  (Metrizable Complete Locally Convex
Topological Vector space).}. The space of positive definite smooth sections of $S^2T^*$ is what we call $\M$. i.e.
$\M=\Gamma^\infty_+(S^2T^*)$, it is a positive  open cone over the vector space $S^2T^*$ (meaning that adding two metrics with positive coefficients is still a metric).

Let us also review the following general facts, which characterize the action of
what will play the role of a Lie algebra and Lie group \cite{Ebin}:
\begin{itemize}
\item The set $\DD:=\diff$ of smooth diffeomorphisms of $M$ is an infinite dimensional Lie group,  and it acts on $\M$
 on the right as a group of transformations by pulling back metrics:
\begin{eqnarray*}
\Psi:\M\times \DD &\ra& \M \\
~(g,f)&\mapsto & f^*g
\end{eqnarray*}
an action which is smooth with respect to the $C^\infty$-structures of  $\M$ and $\DD$\footnote{The natural
action is on the right since of course $(f_1f_2)^*g=f_2^*f_1^*g$.}. We call $\Psi_g:\DD\rightarrow\M$, the action for fixed $g\in\M$, the orbit map. It is clear that two metrics are isometric
if and only if they lie in the same orbit, $$g_1\sim g_2\Leftrightarrow g_1,g_2\in
\mathcal{O}_g:=\Psi_g(\DD)$$
\item The derivative of the orbit map
$\Psi_g: \DD \ra \M$ at the identity is\bea
\alpha_g:=T_{\Id}\Psi_g: \Gamma(TM) &\ra& T_g\M \nn\\
\label{jmath}X&\mapsto & L_Xg \eea where $X$ is the infinitesimal generator of a given curve of
diffeomorphisms of $M$.    The spaces $V_g$, tangent to the orbits will be called vertical and are defined as:
$$ V_g:=T_g(\mathcal{O}_g)=\{L_Xg~|~X\in \Gamma(TM)\}
$$ Since $M$ is compact, every $X\in\Gamma(TM) $ is complete and
$\Gamma(TM)$ forms an infinite dimensional Lie algebra under the usual commutator of vector fields,
$[X_1,X_2]\in\Gamma(TM)$.
\end{itemize}The quotient $\M/\DD$ is known to be a stratified manifold whose singular sets  correspond to the diffeomorphism classes of metrics with non-discrete isometry groups:
$$I_g(M):=\{f\in\DD~|~f^*g=g\}\subset \DD
$$
which are always groups of dimension at most 6. The singular sets are nested according to the dimension of $I_g(M)$.

When dealing with the space of metrics with no symmetries $\M'$, the space $\super'=\M'/\DD$  is indeed a manifold
and the existence of a section \cite{Ebin} allows us to construct
 its local product structure
 $\pi^{-1}(\mathcal{U}_\alpha)\simeq \mathcal{U}_\alpha\times \DD$
through bundle charts for $\mathcal{U}_\alpha$ and open set of the quotient and properly {\it define $\M'$ as a principal fiber bundle} (PFB). With the PFB  $\DD\hookrightarrow \M' \overset{\pi}{\ra}\M'/\DD=\super'$  we have
the usual constructions of gauge theory working properly, as we will see.

  There are other ways to resolve the singularities in the stratified structure of $\M/\DD$ than the one adopted here, which has the disadvantage of excising metrics with high degrees of symmetry such as the ones used to find explicit solutions of the Einstein equations. To excuse ourselves from that obvious criticism, we remark that only a meagre set of initial data will reach such boundaries, that our arguments are of a generic nature and that we can always approximate as well as we like any of those symmetric states. One of the other ways to resolve the singularity involves assuming that the topology of the underlying manifold does not allow for any continuous symmetry group, so called wild topologies, which are infinite in number. Another involves slightly modifying the group $\DD$ one works with, to $\DD_{\{x\}}$ the diffeomorphism which leaves point $x$ fixed. But perhaps the most useful route is to consider not $\M$ but $\M\times F(M)$, where $F(M)$ is the bundle of oriented frames over $M$. Since the action of $\DD$ can be seen to be free over this space, the quotient is indeed a manifold, and it is also a principal fiber bundle over said space. Our view here though is to be minimal with respect to the structures we use.

\subsection{The conformal bundle.}
We now give a brief description of the action of the conformal group $\mathcal{C}$, since it has much nicer
mathematical properties and seems to be given a new importance in recent dual approaches to general relativity
described in the first part of this thesis.

{\flushleft{ \bf Basic results.}}\smallskip

Let $\mathcal{P}$ be the multiplicative group of positive smooth functions on $M$. We denote by
$$\mathcal{C}:=\DD\times\mathcal{P}~~~\mbox{the space of conformal transformations of}~~ M~$$
with group structure $(f_1,p_1)\cdot(f_2,p_2)=(f_1\circ f_2,p_2(p_1(f_2)))$ where $p_2(p_1(f_2))$ just means
scalar multiplication at each $x\in M$ as $p_2(x)(p_1(f_2(x)))$. As with $\DD$, $\mathcal{C}$ is an
infinite-dimensional regular Lie group and it acts on $\M$
 on the right as a group of transformations by:
\begin{eqnarray*}
\xi: \mathcal{C}\times\M &\ra& \M \\
~((f,p),g)&\mapsto & pf^*g
\end{eqnarray*}
For more information on the mathematical properties of conformal superspace and the analogous constructions of
Ebin \cite{Ebin}, see \cite{FiMa77}. For instance, it is fairly easy to prove
in the same fashion as done in section \ref{sec:slice} that a slice theorem exists also for
$\mathcal{C}$ (see  theorem 1.6 in \cite{FiMa77}). Here however, if we want to form properly a principal fiber bundle, we would have to
regard the manifold $\M''$ consisting of metrics with no non-trivial conformal isometries. It can be shown that
this restriction does not have serious topological implications \cite{gomes-2009}.

The derivative of the orbit map $\xi_g: \mathcal{C} \ra \M$ at the identity is given by \bea
\tau_g:=T_{(\Id,1)}\xi_g: \Gamma(TM)\times C^\infty(M) &\ra& T_g\M \nn\\
(X,p')&\mapsto & L_Xg +p'g\eea where $p'\in C^\infty(M)$ and which can be easily evaluated from
$\frac{d}{dt}_{t=0}(tp'+1)(f_t)f_t^*g$, where $f_t=\exp(tX)$.

\subsubsection{York splitting.}\label{sec:York_splitting}

As a last auxiliary result for the main text we here state and sketch the York splitting theorem for the action
of the conformal group of transformations.

From the demonstrated good behavior of the orthogonal projection operators in section \ref{sec:conformal_diff_group}, in all
cases of interest (i.e. for all $\beta$), we have a well defined normal bundle (see \eqref{Splitting alpha}) and
thus a slice for the action of $\mathcal{CD}$. This means that we can write
\be\label{integrable2}
T_g\M=\mbox{Ker}\tau^*\oplus \mbox{Im}\tau_g=T_g\Sigma_g\oplus T_g(\mathcal{CD}\cdot g)
 \ee
where $\Sigma_g$ is the section of $\mathcal{CD}$, given, for the canonical supermetric (DeWitt for $\lambda=0$),
by the exponentiation at $g\in\M$  of the  kernel of $\tau^*$ \eqref{Kernel}. This kernel is composed of
divergenceless (transverse) traceless tensors (TT tensors), which we denote by $S_2^{\mbox{\tiny TT}}\subset S_2T^*$ and can be by
definition decomposed further into
$$S_2^{TT}=S_2^{\DD}\cap S_2^T,
$$
where $S_2^{\DD}\in S_2T^*$ are the transverse (or divergenceless) tensors and $S_T^2\in S_2T^*$ are the traceless
tensors. Exponentiated these spaces respectively form the section $\Sigma_g^{\DD}$ for $\DD$ we used in the subsection \ref{subsec:slice}, and the space of constant volume forms
$$\mathcal{N}_{d\mu}:=\{g\in\M~|~d\mu(g)=d\mu\},$$ where $d\mu(g)$ is the volume form associated to $g$.

Equation \eqref{integrable2} is said to be an \emph{integrable} decomposition in the sense that the tangent space at any point $g$ on the lhs, is the direct sum of tangents to two submanifolds on the rhs.
What is more interesting to us though, is that the second factor in \eqref{integrable2} admits two sets of
integrable decompositions. One of the sets of suborbits, is the natural one given already by the action of the
group written as $\DD\times\mathcal{C}$, as mentioned above. As we are excluding the conformal Killing metrics,
i.e. $\Ker ~\tau=0$, the action of the algebras also clearly splits:
$$ \mbox{Im}(\Gamma(TM))\oplus\mbox{Im}(C^\infty(M))=T_g(\mathcal{CD}\cdot g)
$$
and it is easy to see that the orbit $\mathcal{C}\cdot g$ is a submanifold \cite{FiMa77}, and thus as we already
know this is also true for the $\DD$ group (see section \ref{orbit manifold section}) we have an
integrable decomposition.

The other can be seen by splitting the image of $\tau$ in its traceless and trace part:
\bea\label{York decomposition2} h&=&h_{TT}+L_Xg+Ng=h_{TT}+\frac{1}{3}(2 X^a_{~;a}+3N)g+(L_Xg-\frac{2}{3} X^a_{~;a}g))\\
T_g\M&=&T_g(\Sigma_g)\oplus T_g(\mathcal{CD}\cdot g)=T_g(\Sigma_g)\oplus T_g(\mathcal{C}\cdot g)\oplus
T_g(\mathcal{CD}\cdot g\cap \mathcal{N}_{d\mu(g)})\label{equ:integrable_decomp_conf}
 \eea
For the preceding decomposition to be true in the integrable sense written above \eqref{equ:integrable_decomp_conf}, we only need to check wether
$\mathcal{K}_g:=\mathcal{CD}\cdot g\cap \mathcal{N}_{d\mu(g)}$ is actually a manifold. One can see this since
$$\mathcal{K}_g=d\mu^{-1}(g)\cap \mathcal{CD}\cdot g$$
 where $d\mu:\M\ra \mathcal{V}$ is just the operator that
assigns volume forms to metrics (and $\mathcal{V}$ is the space of volume forms). Thus $\mathcal{K}_g$ is a
manifold, since it is given by an inverse regular value, as can be deduced from $T_gd\mu\cdot h=\text{tr}(h)d\mu(g)$,
which, for $h=L_Xg+Ng$, is surjective. I.e. $\mathcal{K}_g=d\tilde\mu^{-1}(g)$ where
$d\tilde\mu:\M_{\mathcal{CD}\cdot g}=\mathcal{CD}\cdot g\ra \mathcal{V}$. In the end we have the
decomposition: \be\label{York decomposition} T_g\M=T_g(\Sigma^{\DD}_g\cap \mathcal{N}_{d\mu(g)})\oplus
T_g(\mathcal{C}\cdot g)\oplus T_g(\mathcal{CD}\cdot g\cap \mathcal{N}_{d\mu(g)})\ee or in words, we can decompose
the space into ``the volume--form--preserving--divergenceless directions + the scaling--of--the--metric direction + the
diffeomorphisms--that preserve--the--volume--form directions".

\section{Connection Forms}\label{chapter:connection_forms}

In the following we have provided a natural extension of the construction of a principal fiber bundle structure to Riem. That next step is of course, the construction and interpretation of connections for infinite-dimensional groups acting on Riem, and that is the topic of this chapter.

\subsection{Introduction}

It is a much repeated story that in a diffeomorphism invariant theory points lose their meaning, their individuality  becoming
dissolved by the active interpretation of these global ``coordinate changes'' \cite{rovelli}.\footnote{In a sense then our notion of space is nothing but  a $\DD$-torsor (see footnote \ref{footn3}).} In fact, since we will be dealing exclusively with global, and thus active, diffeomorphisms, we will use the expression "\emph{change of labeling}" to distinguish the nomenclature from that of the passive, local ``coordinate changes".

It is the case that in pure gravity
only the metrics over the manifolds attribute any real significance to the spatial points of $M$. We indicate this dependence
by $M_g$, a family of diffeomorphic manifolds parametrized by $g$.

In the canonical analysis, the 3+1 decomposition of the four dimensional metric involves a `shift' vector field and a lapse scalar, which
parametrize the diffeomorphism from a globally hyperbolic space-time to $M\times\R$. This entails in our notation a one-parameter family of diffeomorphic manifolds $M_{g(t)}$.

The  lapse encodes the temporal distance
 element in the embedding of the one parameter family of hypersurfaces.
The 'shift' vector field effectively already requires some
 identification between the points of ``neighboring" $M_{g(t)}$'s.  The shift
 itself is an infinitesimal deviation from the background identification of $M_{g(t)}$ and $M_{g(t+\delta t)}$ by
 vectors orthogonal to $M_{g(t)}$ with respect to the ambient Lorentzian metric. If we propose here to at least
  momentarily disregard four dimensional embedding, specially in view of the first part of this thesis, then the shift vector field loses its meaning, and we must
  find a new way to string together the $M_{g(t)}$'s along time.

The need to somehow identify points of our manifolds along time, naturally brings us to the concept of
best-matching \cite{Barbour94}, and forces us to introduce a form of ``parallel transport" of point labels. From this
concept it is a small step to see this parallel transport as taking place in the gauge setting, where the
structural group should be  $\DD$, since it is this group that parametrizes the ways with which we can
connect $M_{g_0}$ and $M_{g_1}$. It is this concept and its generalization that we will study in this chapter. We regard this construction as an elaboration of the concept of best-matching, as spelled out in \cite{barbour-2002-19}.

Our aim is not just to construct the decomposition $T_g\mathcal{M}=H_g\oplus V_g$ of the principal fiber bundle, but to explicitly construct the Lie algebra valued one form $\omega$. This does not in fact require the introduction of new mathematical apparatus, but it certainly implies a shift in the way one views gauge theory over these configuration spaces towards indeed an original perspective. The connection form then has a very physical interpretation, as something that acts on metric velocities and yields vector fields. Its interpretation is that it yields a preferred infinitesimal ``label change" from each infinitesimal metric change. It is the general study of such connections and what they imply (as for example wrt locality) that this chapter is devoted. In principle, a connection could be derived from a more general action over the whole of Riem(M), including curvature forms. As this requires more work to be made sense of, we start with a certain type of metric induced connections, yielding equation  \eqref{connection from vertical} as our main result.

\begin{figure}
 \begin{center}
 \includegraphics[width=11cm]{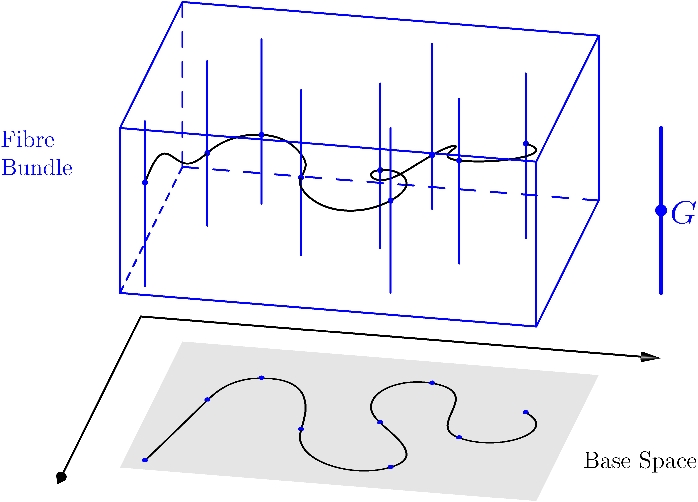}

\caption{A given evolution of the Universe making a trajectory in shape-space, and a given lift to a representation by 3-metrics.  A connection will act by extracting the vertical component of each infinitesimal segment of the line, but it exists on the whole bundle. We'll take it as orthogonally induced by a metric on Riem, but in principle it could be given by other means.}

 \end{center}

\end{figure}

\subsection{Connection forms: basic properties}

Let us first of all define a connection form for a finite-dimensional principal fiber bundle $P$:
\begin{defi}
 In finite-dimensions a connection form is defined as a Lie-algebra-valued one form $\omega\in\Gamma(T^*P\otimes\Lg)$ which acts on vertical vectors as: $\omega(T_{\Id}\mu(X))=X$, where $X\in \Lg$ and $\mu:P\times G\rightarrow P$ is the group action. The connection must also transform as
  \end{defi}
 $$ R_h\omega=\Ad(h^{-1})\omega
 $$
In the above $R_h$ is the push-forward right action: $R_h\omega(v)=\omega((\mu_h)_*v)$ where $v\in T_pP$ and $\mu_h:P\rightarrow P$ is the right action by the group element $h$. The adjoint acts on the algebra as
$$\Ad(h)X=\frac{d}{dt}_{|t=0}\left(h\circ \exp(tX) h^{-1}\right)
$$ where $\exp:\Lg\rightarrow G$ is the group exponential. Let us also define \emph{equivariance} as a term that encompasses both the covariant and contravariant denominations. Different spaces might have different equivariance properties. For instance the usual Yang-Mills curvature form transforms equivariantly in the adjoint representation, but the local expression for the connection does not, as it does not transform homogeneously.

The connection form in our present infinite-dimensional  setting will then be a Lie-algebra valued linear functional on $T\M$ (metric velocities).
 Since technically  the Lie algebra here is just the space of infinitesimal diffeomorphisms of $M$, the
connection form turns out to be a vector-field-valued distribution, taking metric velocities as test functions. We
are thus led to the meaning of a gauge connection over $\M'$ as representing a Machian notion of relational space,
since it relates spatial points along time in a manner depending on the dynamics of the entire
Universe (depending strictly and globally on the metric velocities). The connection form is not an empty mathematical construct. Its interpretation as yielding parallel transportation of spatial points\footnote{The connection form for the conformal group would then yield a notion of parallel transport of local scale.} is suggestive and interesting (especially from a relational point of view).

\subsection{ An equivariant splitting.}

  A choice of connection form in $P$ amounts as usual to choosing an
 equivariant decomposition
 \be\label{equivariant decomposition}T_g\M'=H_g\oplus V_g.\ee
 where by equivariant we mean that the decomposition is maintained by the group action.

   In the infinite-dimensional case, we have to separate the requirement of the direct sum in three separate conditions:
 \be\label{connection conditions} \begin{array}{c} V_g\mbox{~~and~~}H_g\mbox{~~are closed~~}\\
 V_g\cap H_g=\{0\}\\
 \overline{\mbox{span}\{V_g\cup H_g\}}=T_g\M'
 \end{array}\ee
 We will shortly discuss the mathematical instruments and conditions required for  these conditions.
  The physical purpose of the decomposition is so that we can distinguish what is a label change, which would be given by projection of  metric change along the orbits, and the remnant of the metric change, which could be identified with pure geometrical change.

 This amounts to defining
 an equivariant projection on the vertical space, which we call $\hat \V:TP\ra V$,
 having the properties $\hat \V\circ\hat \V=\hat \V$ and equivariance: $f^*\circ \hat\V=\hat \V\circ f^*$. But  to be able to construct the actual connection form we need more than just the equivariant direct sum decomposition, but also conditions on the representation of the algebra (that it forms a fundamental vector field with the correct equivariance properties). This is necessary to derive the correct transformation properties of the connection itself $R_f\omega=\Ad(f^{-1}) \omega$. We also include this criterion and show that indeed such a connection has these properties.

 \subsection{ The connection form obtained from the vertical projection}

The vertical sub-bundle, the bundle tangent to the orbits, is given by
$V_g:=\{L_Xg~|~X\in \Gamma(TM)\}$, where $L_X$ is the Lie derivative.
The
canonical  representation of  the diffeomorphism group on $\Gamma(TM)$ is the adjoint representation:
$$
 \Ad(f)(X)= \frac{d}{dt}(f\circ \exp(tX)\circ f^{-1})=f_*(X\circ f^{-1}).
$$
 Now, for the representation of the Lie algebra  on itself, we have
 $\ad_X=[X,\cdot]$ since $[X,Y]=\frac{d}{dt}_{|t=0}\Ad(\exp(tX))Y~~\mbox{which is an element of}~~\Gamma(TM).$
  Hence the Lie algebra bracket is just the vector field commutator.

  If we have a right action of a Lie group on any manifold $\mathcal{N}$ ($\mathcal{N}$ is allowed to be infinite-dimensional) $\Psi:\mathcal{N}\times G\ra \mathcal{N}$,
  for $X\in\Lg$ we define the {\it fundamental vector field} $\zeta_X\in\Gamma(T\mathcal{N})$
  by
  \be\label{equ:fundamental_vf}\zeta_X(x)=(T_{(x,\Id)}\Psi)\cdot(0_x,X)=:T_{\Id}\Psi_x(X),\ee
  where we redundantly keep the subscript $x$ in $O_x$, to remind ourselves that this is the zero vector at the point $x\in P$.
  \begin{lem}\label{integrable}
  A fundamental vector field must satisfy the following properties for $f\in G$:
  \begin{enumerate}
\item $\zeta:\Lg\ra\Gamma(T\mathcal{N})$ is linear.
\item $T_x(\Psi_f)(\zeta_X(x))=\zeta_{\Ad(f^{-1})X}(\Psi(x,f))$
\item $[\zeta_X,\zeta_Y]=\zeta_{[X,Y]}$
\end{enumerate}\end{lem}
where we have denoted, for fixed $f$, in the same way as in \eqref{equ:fundamental_vf}:
\be\label{equ:push_forward_at_f} T_x(\Psi_f):=(T_{(x,f)}\Psi)\cdot(\cdot, 0_{f}): T_x\mathcal{N}\rightarrow T_{\Psi(x,f)}\mathcal{N}
\ee

Let us explicitly check property 2 for the $\DD$ action on $\M$. To make the
actions clearer we expand $f^*g=\Psi(f, g)$ and so we have that, according to \eqref{equ:push_forward_at_f}, for $f\in\DD$ and $g\in\M$,
 \be\label{equ:fundamental_vf_diff}T_g(\Psi_{f}):=(T_{(g,f)}\Psi)\cdot (\cdot,0_f)
 :T_g\mathcal{\M}\ra
T_{\Psi({f},g)}\mathcal{\M}
\ee and
\be\label{equ:push_forward_at_f_diff} T_{\Id}\Psi_g:=(T_{(g,\Id)}\Psi)(0_g,\cdot ):\Gamma(TM)\ra T_g\M.
\ee
Using $r_f$ for the right action of the group on itself we also have:
 \be\label{equ:conjugate_action}\Psi(g,  r_f\circ h)=\Psi(f^*g,\mathrm{conj}_{f^{-1}}(h))\ee for any $h\in\DD$,
 where $\mathrm{conj}_f$
 is the conjugate action of the group on itself: $\mathrm{conj}_f(h)=f\circ h\circ f^{-1}$, and again $\circ$ is composition of maps (in this case diffeomorphisms).

Using \eqref{equ:push_forward_at_f_diff} and \eqref{equ:fundamental_vf_diff} we have: \bea
 T_g(\Psi_f)(L_X(g))&=& (T_{(g,f)}\Psi)\cdot ((T_{(g,\Id)}\Psi)\cdot(0_g,X ),0_f)\nn\\
 &=&(T_{(g,f)}\Psi)\cdot \left((T_{(g,\Id)}\Psi)\cdot(0_g,\frac{d}{dt}_{|t=0}\exp (tX) ),0_f\right)\nn\\
 &=&(T_{(g,f)}\Psi)\cdot \left(\frac{d}{dt}_{|t=0}\Psi(g,\exp(tX)),0_f\right)\nn\\
 &=&\frac{d}{dt}_{|t=0}\left(\Psi\left(\Psi(g,\exp(tX)),f\right)\right)=
 \frac{d}{dt}_{|t=0}\left(\Psi(g,\exp(tX)\circ f)\right)\nn\\
  &=&\frac{d}{dt}_{|t=0}\left(\Psi(f^*g,\mathrm{conj}_{f^{-1}}\exp(tX))\right)=
  \frac{d}{dt}_{|t=0}\left(\Psi(f^*g,\exp(t\Ad_{f^{-1}}X))\right)\nn\\
\label{omegaprop} ~&=&L_{\Ad(f^{-1})X}(f^*g),
 \eea where we used \eqref{equ:conjugate_action} to go from the fifth to the sixth line.

  We can then identify $\zeta_X$ in {\bf Lemma \ref{integrable}}
  with $L_X$ and verify that it automatically satisfies the first and third
  identities required for  {\it a fundamental vector field}.
 So the key properties of the action of the
  Lie algebra on the bundle are satisfied by the Lie derivative of vector fields, a fact that is of utmost importance for our treatment and that allows us to take the ``gauge analogy" to be not merely an analogy.

   We then
  define the Lie-algebra valued connection form:
  \begin{defi}
  Given a tangential decomposition as in \eqref{connection conditions}, if we can construct a vertical projection $\hat \V:TP\ra V$ satisfying $\hat \V\circ\hat \V=\hat \V$ and\footnote{ We pause to note that for usual push-forward and pull-back maps for ``constant" diffeomorphisms, we could have used
extra ``$*$"'s: $(f^*)_*:T_g\M\ra T_{f^*g} \M$. This is the exact analogous of the tangent left translation
${l_g}_*$ in the usual action of Lie groups. In this setting this is superfluous since due to the vector space
structure of $S_2M$, we have  ${f^*}_*=f^*$. From now on we will omit the double star notation.} $f^*\circ \hat\V=\hat \V\circ f^*$, we then define the vector field valued connection form as:
\be\omega_g=(T_{\Id}\Psi)^{-1}_g\circ \hat V_g:T\M'\ra\Gamma(TM)\ee
\end{defi}
Since
$\alpha_g:=T_{\Id}\Psi_g$ is an isomorphism over its image, i.e. over the vertical space, we can take
inverses.

Clearly we then have that
\be\label{vertical connection}\hat\V_g[\dot g]=\alpha_g\circ\omega_g[\dot g]=L_{\omega_g[\dot g]}g\ee where
$\dot g\in T_g\M$. Thus vertical projection of a velocity is equal to the Lie derivative of the metric along the preferred direction $\omega[\dot g]$.

By the transformation properties of the Lie derivative (see \eqref{omegaprop}) and equivariance of the vertical projection we have the usual
transformation property: \be\label{connectionprop} f^*\omega=\Ad(f^{-1})\omega \ee
confirming that this can indeed be interpreted as a connection form.

Because $\DD$  admits the exponential map, we have not only uniqueness, but also
existence of a $\DD$-equivariant smooth parallel transport. This means that we can integrate forward from some initial labeling, finding some relative preferred identification of spatial point along time.\footnote{We do note for completeness that it is well known that the exponential map of $\DD$ is not surjective on any neighborhood of the identity. We however abstain from speculating on the relevance of this fact for our approach.} To be more explicit, let us consider a heuristic example. Let us suppose we describe (supposing we could observe the entire Universe) a time evolution of the metric, i.e. a one-parameter family $g(t)$. Given an initial labeling of the manifold at $t=0$, we can then integrate the connection along time to find the preferred point that is ``equilocal"  to $x(0)$:
\be \label{equ:integral_bm}x(0)\mapsto \int_{[0,1]}\exp{\omega[\dot g(t)]}( x) dt.
\ee

 \subsubsection{Interlude: interpretation of non-trivial holonomy.}
 So what would it mean to have two paths in shape space $[g_1(t)]$ and $[g_2(t)]$ (this hypothetical situation then by definition already falls outside of the domain of classical physics), starting from the same shape $[g_0]$, such that their horizontal lifts $g^H_1(t)$ and $g^H_2(t)$, although arriving at the same shape $[g^H_1(1)]=[g^H_2(1)]$, fall on different places on the orbit $g^H_1(1)\neq g^H_2(1)$? For the diffeomorphism group, it would mean that best matched observers who agree on the ``location" of points initially would, through \eqref{equ:integral_bm}, disagree on their location in the end. We leave a thought her for the reader: what would it mean for the conformal group?

\subsection{ Locality of connection forms.}

Another factor of extreme importance  is the question of local representability of the connection form. That
is, $\omega$ at each $g$ is an element of $T^*_g\M\otimes\Gamma(TM)$. However, since we are dealing with
infinite-dimensional spaces, we cannot a priori identify the space of linear functionals acting on
$T_g\M'\simeq\Gamma(S_2T^*)$, which we call $T^*_g\M'$,  with $\Gamma(TM\otimes_STM)=\Gamma(S_2T)$.

As an initial attempt to construct such a local representation, we could choose a partition of unity of $M$,
defined by the characteristic functions $\{\chi_\alpha\}$ of the open sets $\{U_\alpha\in M\}$. Then for an
element $\lambda_g\in T_g\M'$ by linearity we have:
$$\lambda_g[\dot g]=\sum_\alpha{\lambda_g}_{|_{U_\alpha}}[\dot g_{|_{U_\alpha}}]
$$
where in this section we have denoted functional dependence by square brackets. In the limit,  this would come to: \be\label{dense representation}
\lambda_g[\dot g]=\int_M\lambda_g^{ab}(x)\dot g_{ab}(x)d\mu_g \ee for $\lambda_g\in\Gamma(TM\otimes_STM)$.

In fact, what we have is that elements of $T_g\M$ are tensors with compact support, which can thus be considered
as a space of {\it test functions} (or more precisely, test tensors\footnote{Since $M$ is compact, we can take the
components of an element of $T_g\M$ as the test functions.}). We have that the space $T^*_g\M'$ is of course {\it a space of
distributions on $T_g\M$}, and the space defined above by elements of the simple form of \eqref{dense representation} are dense inside $T^*_g\M$.

We will ignore these subtleties and now express $\omega$  (up to discrepancies on sets with vanishing measure) as the two-point tensor:
$\tilde\omega\in\Gamma(TM)\otimes\Gamma(TM\otimes_S TM)$. Pointwise:
 \begin{eqnarray}
\omega_g(x,x') ~
 \in (T_{x'}M\otimes_S T_{x'}M)\otimes T_xM\simeq L(T^*_{x'}M\otimes_S T^*_{x'}M, T_xM) \nn\\
  \label{connection} \int \omega^{ab'c'}(\dot
g_{b'c'}(x')) \sqrt{g}d^3x'=\omega^a_g[\dot g](x)\in T_xM\end{eqnarray} where we have used DeWitt's notation,
denoting tensorial character at $x'$ by primed indices. In the examples we will find, the connection form will always be given by the simple form of \eqref{dense representation}.

 The geometrical interpretation of the connection form viewed in this way is that, for each metric $g$, a
given metric velocity $\dot g(x)$ at a point $x\in M$ will contribute for the ``best-matching'' vector field at
each other point $x\in M$. In this way then, we get a non-local contribution to the best-matching vector field at
each point of $M$. These contributions however may  come from metric velocities at that and every other point of
$M$. This goes in line with relational arguments, since this implies that the stringing of points throughout time (equilocality) are determined by the \emph{kinematics of the entire universe}.

\subsection{Construction of connection forms in $\M$ through orthogonality}\label{Explicit construction}

Now that we have written down the basic structures that allow a gauge treatment of labelings using the $\DD$
group, we will derive explicit formulae for the connection forms through the use of the orbit maps and their adjoints. Of course in this
case a supermetric fixes the connection, if it exists, once and for all. It is not, as it is in Yang-Mills, determined by an action principle. So when we consider actions involving such fixed connections,
 our system works in analogy to a particle in a fixed electromagnetic potential.

 To let the connection be determined through a variational principle, one would require a term like $F[\Omega]$ in the action, where $\Omega$ is the curvature form of $\omega$. However, as $\Omega[\dot g,\dot g]=0$, it
 cannot appear in the `classical trajectory in $\M$' action we are considering.
 Nonetheless this can be done for a field theory in $\M$, a treatment which will show up in future work.

 Now we shall address the three items in \eqref{connection conditions} and present a direct formula for the connection form and the pre-requisite conditions on its components. We emphasize that no mathematical breakthrough is needed in  the construction of the vertical projection The only thing that is mathematically novel in the present work is the use of a connection, and  this is a very direct consequence of $i)$ $\Gamma(TM)$ forming a Lie algebra and $ii)$ the Lie derivative of the metric along these fields forming a fundamental vector field on $T\M$.

 Nonetheless, the emphasis here is completely different than that of usual treatments, as it introduces the explicit use of a connection form and focuses on the conditions necessary for its definition  from the basic ingredients. We will  generalize the statements in this section to the case of the conformal group.

\subsection{Construction through orthogonality and the momentum constraint}\label{sec:orthogonal_momenta_diff}

We could initially attempt to define an equivariant direct sum decomposition \eqref{equivariant decomposition} implicitly, through an equivariant inner product in $\M'$, by orthogonality with respect to the vertical bundle.

In other words, by  defining the horizontal subspace $\HH$ by orthogonality to the canonical fibers with respect to some
$\DD$-invariant supermetric $\langle\cdot,\cdot\rangle$: \be\label{ortho}
\mathcal{G}[\hat H[\dot g], L_Zg]=\int_M\langle \dot g-\hat V_g[\dot g], L_Zg\rangle dx^3=\int_M\langle\dot
g-L_{\omega[\dot g]}g, L_Zg\rangle_gdx^3=0 \ee
For instance, writing the canonical momentum as:
$$\pi^{ab}=\frac{1}{2N}G^{abcd}(\dot g_{cd}-L_\omega g_{cd})
$$ we obtain that the momentum constraint is written as
\be\label{momentum constraint}\mathcal{H}^a=\left(\frac{1}{2N}G^{abcd}(\dot g_{cd}-L_\omega g_{cd})\right)_{;b}=0\ee
 For $\omega=\omega_g[\dot g]$ given by \eqref{vertical connection}, we have that \eqref{momentum constraint} is implicitly exactly of the form of \eqref{ortho}, with $\langle\cdot,\cdot\rangle=\frac{G^{abcd}}{N}$:
$$\int_M\frac{1}{2N}G^{abcd}(\dot
g_{ab}-L_{\omega[\dot g]}g_{ab} )L_Zg_{cd}dx^3=0
$$ which has to be valid for all $Z\in\Gamma(TM)$. This is how the momentum constraint can be shown to be merely a statement to the effect that  the connection is induced by orthogonality to the fibers.

 If a horizontal space  is well defined with respect to such an invariant
supermetric, the projections should  themselves be equivariant, e.g. $(f^*)^*\hat \V_g=\hat\V_{f^*g}\circ{(f^*)}_*$
 However, the fact  that one is dealing with (completions in) function spaces obstructs such a direct approach. In the infinite-dimensional setting one cannot know if for instance $H$ and $V$ are closed, the first requirement of \eqref{connection conditions}. Furthermore, this procedure does not provide us with an explicit formula for our connection form. Even the vertical projection operator, which would in the finite dimensional case be defined as $P(w)=\sum_i\langle v_i, w\rangle v_i$, for $v_i$ an orthonormal basis for $V$, requires modification in the present infinite-dimensional case.

 We shall proceed differently, and find that there exists a more comprehensive way to define
a vertical projection operator and valid connection explicitly. This includes the orthogonality criterion. For this we need to
use the Fredholm alternative.

\subsection{Using the Fredholm Alternative}

In Subsection \ref{projection section} we have shown that  if the horizontal bundle is defined as the space
orthogonal to the orbits, i.e. orthogonal to $\mbox{Im}(\alpha)$ (where we remind the reader that $\alpha$ is the tangent to the orbit map at the identity), $H$ is given by $\mbox{Ker}(\alpha^*)$, (since $(\alpha(X),v)=0$
if $v\in\mbox{Ker}(\alpha^*)$).

 Taking any supermetric, without further assumptions
 (see alternative formulation of Theorem \ref{proj theo}):
 \begin{itemize}
 \item  The operator $\alpha$ and also its symbol $\sigma(\alpha)$ are injective. The first of these requirements is equivalent to a restriction to configurations that do not possess symmetry wrt the relevant group. For $\DD$ for example, it amounts to restricting our attention to $\M'$;
 \item A smooth adjoint of $\alpha$ exists
 with respect to the fiber metrics in $TM$ and $T^*M\otimes_S T^*M$,
 such that $\Ker(\alpha^*)\cap\mbox{Im}(\alpha)=0$ and $\Ker(\sigma(\alpha^*))\cap\mbox{Im}\sigma(\alpha)=0$. The condition $\Ker(\alpha^*)\cap\mbox{Im}(\alpha)=0$ can be seen in fact to be equivalent to requiring the supermetric on $\mbox{Im}(\alpha)$ to be non-degenerate (see  Proposition \ref{propo} in section \ref{sec:constructingV}). This is a condition automatically implemented in the  Fredholm alternative that encodes the criterion for $H_g\cap V_g=0$;
 \item The supermetric is $\DD$-equivariant. \end{itemize}
 Then the  operator defined by \eqref{P vertical}: \be \hat
V:=\alpha\circ(\alpha^*\circ\alpha)^{-1}\circ\alpha^*:T\M'\ra T\M'\ee is well-defined and satisfies all required
properties of a vertical projection operator: \begin{itemize}\item $\hat V$ is $\DD$-equivariant. \item It is idempotent,
$\hat V^2=\hat V$.
\item $\hat V(\alpha(X))=\alpha(X)$ for $X\in\Gamma(TM)$.
\item The space orthogonal to the orbits (or horizontal) satisfies:
$H:=\nu(\mathcal{O}_g)_h=\mbox{Ker}(\alpha_h^*)=\mbox{Ker}\hat V_h$ and $V=\mbox{Im}(\hat V)$ and thus
$T_g\M=H\oplus V$.
\end{itemize}
 In fact, the invariance of the supermetric
 is only used in the construction of the $\hat V$ operator in order to find  the necessary
transformation properties of $\alpha^*$. It ensures that the adjoint of $(r_f^{-1})_*$, where $r_f$ is right
translation by $f$,  is indeed $(r_f)_*$ and so $\alpha^*\circ\alpha$ transforms in the appropriate way. It is also worth noticing that the appearance of the inverse differential operator $(\alpha^*\circ\alpha)^{-1}$ in the definition of the vertical projection operator confirms the non-locality of the connection form explicit in \eqref{connection}.

From the vertical projection operator we obtain the connection form in the usual way:
\be \label{connection from vertical}
\omega:=\alpha^{-1}\circ\hat V=(\alpha^*\circ\alpha)^{-1}\circ\alpha^*\ee
Note that if the vertical operator is well-defined, so is $\alpha^{-1}_{|V_g}$.

\subsection{Equivariant metrics}

 We here first list a wide range of  inner products in $\M$ which are
$\DD$-invariant. We are able to prove equivariance for any supermetric of the form $FG_\beta$ where
$G^{abcd}_\beta=g^{ac}g^{bd}-\beta g^{ab}g^{cd}$  is a one-parameter family of supermetrics,
 weighted by a functional $F:\M\ra C^\infty(M)$, which we call {\it lapse
potential} and define as: \begin{defi}\label{lapse potential} A lapse potential is any functional $F:\M\ra
C^\infty(M)$ formed from $g$ and its curvature tensor by means of tensor product, index raising or lowering,
contraction and covariant differentiation.\end{defi}

To prove that the above mentioned class of supermetrics indeed induces an invariant inner product,  one must simply apply
 a theorem (see for instance {Theorem 9.12.13} of \cite{Bleecker}) which establishes  that, for such a lapse potential $F$,
$F(f^*g)=F(g)\circ f$. Furthermore it is easy to show that $L_Zg$, for any $Z\in\Gamma(TM)$, is a Killing vector
for the generalized supermetric  \cite{Giulini:1993ct}. Combining these facts we have:
\be \int_M \frac{1}{F(f^*g)}G_\beta(f^*u,f^*v)_{f^*g}d\mu_{f^*g}=\int_M (\frac{1}{F(g)}G_\beta(u,v)_{g}\circ
f)f^*d\mu_g.\ee

\subsection{ Ellipticity of $\alpha^*\circ\alpha$}

We have already shown  that $\alpha$ has injective symbol in subsection \ref{orbit manifold section}, furthermore, by the very definition of $\M'$, it obviously true that
$\alpha$  is injective over $\M'$.
\begin{prop}\label{prop:injective_symbol} For each $g\in \M'$, for the inner products $g$ and
$G_{\beta}/N$ in $TM$ and $T^*M\otimes_{\mbox{\tiny S}}T^*M$ respectively, for $\beta\neq 1$ and $N$ any lapse
potential, $\Ker(\sigma(\alpha^*))\cap\mbox{Im}(\sigma(\alpha))=\emptyset$.
\end{prop}
{\rm Proof.~ } We first calculate the symbol of $\alpha^*$ (see subsection \ref{orbit manifold section}).
 For $\lambda\in T_x^*M$ and $v\in T_xM$
 such that $\xi=g(v,\cdot)$, we have $\sigma_\lambda(\alpha):T_xM\ra T_x^*M\otimes_{\mbox{\tiny S}}T_x^*M$ given by
 \be\label{equ:symbol1} \sigma_\lambda(\alpha)(v)=\xi\otimes_{\mbox{\tiny S}} \lambda=2v_{(a}\lambda_{b)}
 \ee From now on we omit the $\alpha$ in the notation. For $u_{ab}\in T_x^*M\otimes_{\mbox{\tiny S}}T_x^*M$
   the adjoint symbol can be directly defined by:
  $$\frac{G^{abcd}_\beta}{N} u_{ab}(\sigma_\lambda(v))_{cd}=(\sigma^*_\lambda(u))^cv_c$$  From this one easily
  calculates (we also omit the $\beta$ dependence to avoid cumbersome notation)
  $\sigma^*_\lambda: T_x^*M\otimes_{\mbox{\tiny S}}T_x^*M\ra T_xM$:
  \be(\sigma^*_\lambda(u))^a= \frac{2}{N(x)}(u^{(ab)}\lambda_b-\beta u^a_{~a}\lambda^a)\label{adjoint symbol}
  \ee
Now inserting $u_{ab}=\sigma_\lambda(v)=2v_{(a}\lambda_{b)}$ for some $v$, and assuming $u_{ab}\in\Ker(\sigma^*_\lambda)$,
we have \bea u^{(ab)}\lambda_b&=&\beta u^a_{~a}\lambda^a\nn\\
\frac{1}{2}||\lambda||^2v^a&=&(\beta-\frac{1}{2})( v^b\lambda_b)\lambda^a\nn
 \eea and thus $\lambda^a=c v^a$, which fed back into the equations can easily be seen to only
 have a solution for $\beta=1$.
$\square$.

So if we are not approaching $\beta=1$ this part of the requirements for the vertical projection operator for such supermetrics is satisfied. However, the value $\beta=1$ is the one present in the canonical 3+1 decomposition of general relativity, and is the one value for which one retains foliation invariance. However, there exist different approaches to gravity that encode fixed foliations, such as \cite{Horava:lif_point} and the theory of Shape Dynamics presented in the first part of this thesis. In particular, SD yields a theory even classically dynamically equivalent to GR which does include a preferred foliation through the use of the conformal group as a symmetry group.

As we will see, a trivial consequence of using the conformal group is that  the equivalent of
$\alpha^*\circ\alpha$ is indeed elliptic for all $\beta$ and lapse potentials.


\subsection{The intersection $\Ker(\alpha^*)\cap\mbox{Im}(\alpha)$}\label{Intersection section}

 There is a  potential problem even in the simple implicit
  orthogonality view, which stems from the non-definiteness  of the deWitt supermetric.
 If the direct sum decomposition is to be
determined by an orthogonality relation with respect to a metric that is not definite (it has signature
$-+++++$), we could run the risk of having elements of the vertical space that are orthogonal to the vertical
space, i.e. $v\perp\V_g$ such that $v=L_Xg$ for some $X\in\Gamma(TM)$, hence $v\in\V_g$ as well.

The adjoint $\alpha^*$ is given by:
\be\int_M
\frac{1}{N}G^{abcd}_\beta u_{ab}X_{(c;d)} d\mu_g=\left(\frac{1}{N}(u^{cd}-\beta g^{cd}u^a_{\phantom{a};a})\right)_{;c}X_d
\ee
Thus
\be \alpha^*(u_{cd})=\left(\frac{1}{N}(u^{cd}-\beta g^{cd} u^a_{\phantom{a};a})\right)_{;c}
\ee
However, even if we simplify the treatment to the case where the functional
$N(x;g]=N[g]$ is spatially constant, we can already glimpse severe obstructions to $\Ker(\alpha^*)\cap\mbox{Im}(\alpha)$ having zero  intersection. First note that
\begin{eqnarray*}
g^{ac}{X}_{(a;b);c}&=&\frac{1}{2}g^{ac}({X}_{a;bc}+{X}_{b;ac})=\frac{1}{2}g^{ac}(R^d_{~abc}{X}_d+{X}_{a;cb}+{X}_{b;ac})\\
~&=&\frac{1}{2}(R^d_{~b}{X}_d+({X}^d_{~;d})_{;b}+\nabla^2{X}_b)
\end{eqnarray*}
where
$\nabla^2{X}_b:=g^{ac}({X}_b)_{;ac}$ is the Riemannian Laplacian. Then
\be\label{one}\alpha^*\circ\alpha(X)= (g^{ac}g^{bd}-\beta g^{ab}g^{cd})({X}_{(a;b)})_{;c}=
\frac{1}{2}(R^{db}{X}_d+(1-2\beta)({X}^d_{~;d})^{;b}+\nabla^2{X}^b) \ee
 If one assumes $\beta\neq 1$, then the operator is elliptic. However, even for $\beta=1$, it can be shown  that   non-trivial \footnote{Since we have excluded Killing fields from our considerations, trivial
solutions to these equations are the ones for which $X_a=0$.} sets of solutions (or lack
thereof) of \eqref{one} (which is equivalent in this case to $\Ker(\alpha^*)\cap\mbox{Im}(\alpha)=0$) depend on the metrics $g$ \cite{Giulini:1993ct}.
There are a number of solutions and domains of validity for the condition
$\HH_g\cap\V_g=\{0\}$ even for $\beta=1$. For example,  for all Ricci-negative geometries (which always exist for
closed $M$ \cite{1986InMat..85..637G}) the condition holds, as well as for non-flat Einstein metrics. For a more
extensive study of these matters see \cite{Giulini:1993ct}. We remark though that as this would still not give a splitting of $\M'$, it would not count as a connection in the sense applied here, which requires it to exist on the whole principal fiber bundle.

\section{The conformal diffeomorphism group.}\label{sec:conformal_diff_group}

Now we apply the same reasoning as in the previous section to the case of the conformal group.

The symbol of $\tau_g$, for $\lambda\in T_x^*M, v\in T_xM$ and $c\in\R$ can be seen to be
 \be\label{symbol tau}
\sigma_\lambda(v,c)=cg_{ab}+\lambda_{(a}v_{b)} \ee Now,  take the metric $\langle\cdot,\cdot\rangle$ in
$T_x^*M\otimes_{\mbox{\tiny S}}T_x^*M$ to be $NG_\beta$. \footnote{Note that in our notation the lapse potential
here appears multiplying the metric, as opposed to the usual lapse in ADM which for the kinetic term appears
dividing it. This will make it easier to deal with powers and negative signs. } The inner product in
$T_xM\times\R$ is taken to be $g(v_1,v_2)+c_1c_2$. Then for $u_{ab}\in T_x^*M\otimes_{\mbox{\tiny S}}T_x^*M$
  from the definition of the adjoint symbol:
  $$NG^{abde}_\beta u_{ab}(\sigma_\lambda(v,c))_{de}=\langle(\sigma^*_\lambda(u)),(v,c)\rangle$$ we easily
  find
  $\sigma^*_\lambda: T_x^*M\otimes_{\mbox{\tiny S}}T_x^*M\ra T_xM$
  \be(\sigma^*_\lambda(u))=
  \left(2(u^{(ab)}\lambda_b-\beta u^a_{~a}\lambda^a),-(1-3\beta)u^a_{~a}\right).\label{adjoint symbol conformal}
  \ee
{\flushleft{ \bf Ellipticity of $\tau^*\circ\tau$}}\smallskip

 Now suppose $u_{(ab)}=\sigma_\lambda(v,c))_{ab}$
and $(\sigma^*_\lambda(u))=(0,0)$. Then we have that
\be\label{equ:symbol_conf}\begin{array}{rcl} c\lambda^a+||\lambda||^2v^a+\lambda^bv_b\lambda^a&=&0\\
3c+2\lambda^av_a&=&0\\
\Rightarrow ||\lambda||^2v^a-\frac{c}{2}\lambda^a&=&0\end{array}\ee
 Contracting the last equation with  $\lambda_a$  yields $-2||\lambda^2||c=0$ which only has solution for $c=0$, in which case
$v^a$ is also obligatorily zero as well. Thus we have proven that
\begin{prop}
For the given action of $\mathcal{C}$ on $\M$, $\alpha$ is an elliptic  operator and
$\Ker(\sigma(\tau^*))\cap\mbox{Im}(\sigma(\tau))=0$. Thus $\tau^*\tau$ is an elliptic operator. \end{prop}

{\flushleft{ \bf The intersection $\Ker(\tau^*)\cap\mbox{Im}(\tau)$}}\smallskip

Since we have gone directly to the calculation of the symbol $\sigma^*(\tau)$, we now write down the actual
operator, for $v_{ab}\in \Gamma(S_2T^*)$. First of all, we check that the supermetric defined by
\eqref{supermetric} is equivariant with respect to the action of $\xi$ (global gauge transformations). One must
merely see that $\xi(f,p)$ acts on the covariant metric tensor $g^{ab}$, as
$$ \xi((f,p),g^{ab})=p^{-1}f^{-1}_*g^{ab}
$$
Thus for global transformations we have:
\be\mathcal{G}[u,v]=\int_Md^3x
\sqrt{pf^*g}N_{pf^*g}G^{abcd}_{pf^*g}(pf^*u)_{ab}(pf^*v)_{ab}=\int_Md^3x\sqrt{g}N_gG^{abcd}u_{ab}v_{ab} \ee  where
for the supermetric to be conformally invariant, $N$ must now not only be a {\it lapse potential}, but also must be further constrained:
 \begin{defi}\label{defi:conf_lapse}
A conformal lapse functional is one for which  $N_g(x)>0$ and
 \be\label{lapse potential
2}N_{pf^*g}(x)=p^{-3/2}N_g(f(x))\ee
\end{defi} We will give an example of
such a lapse potential below.

Calculating the adjoint operator we get: \be\label{Kernel} \tau^*(v)=\left(-2Nv^a_{~a}~,-(NG^{abde}_\beta
v_{de})_{;b}\right) \ee Since for the kernel of $\tau_\beta^*$ the trace part of $v_{ab}$ is zero from the first
component of \eqref{Kernel}, we immediately see that  (inputting the $\beta$ back into the notation for the
adjoint) $G^{abde}_\beta v_{de}=G^{abde}_0v_{de}$. Thus $\mbox{Ker}\tau^*_{~\beta}=\mbox{Ker}\tau^*_{~0}$. Hence
$\Ker(\tau^*_{~\beta}\circ\tau)= \Ker(\tau^*_{~0}\circ \tau )$. Thus, under the supposition that the lapse is a
strictly positive function, exactly as the result $\Ker(\alpha^*_{~0})\cap\mbox{Im}(\alpha)=0$ contained in Section \ref{sec:slice} (see also \cite{FiMa77}), a result dependent on a positive definite inner product on both the target and
domain spaces, one can show that $\Ker(\tau^*_{~0})\cap\mbox{Im}(\tau)=0$. For completeness, the specific
equations for elements of $\Ker(\tau^*_{~0})\cap\mbox{Im}(\tau)$, are, for $v_{ab}=X_{(a;b)}+\lambda g_{ab}$
\be\label{conformal symbol}
\begin{array}{c}
X^a_{~;a}+3\lambda=0\\
\left({N}(X^{(c;d)}+\lambda g^{cd})\right)_{;d}=0\end{array}\ee

 {
\flushleft{ \bf Equivariance of $\tau^*\circ\tau$.}}\smallskip

 Now all that is left to prove that indeed we
have a well-defined connection form for the conformal group (given implicitly by the generalized metrics in $\M$)
is to check wether $\hat V$ transforms equivariantly, or in other words, that  $\tau^*\circ\tau$ is extended to be
right invariant. As one can see from  equation \eqref{equivariant V}, this is dependent strictly on equation
\eqref{orbit transf}, which we now compute for this action. This computation is equivalent to finding out if the action of the group produces a fundamental vector field, as we did for the diffeomorphism group.

The left hand side of \eqref{orbit transf} gives, for $f_t$ the integral
diffeomorphism of $X$: \be T_{(f_0,p_0)}\xi_g(X\circ f_0,p'\circ
f_0)=\frac{d}{dt}_{|t=0}\xi((f_t,p_t),g)=f_0^*(p'g+p_0L_Xg) \ee where $f_t$ produces the integral curves of the
field $X(x)=\frac{d}{dt}_{|t=0}f_t(f_0(x))$. In its turn the right hand side  gives:
 \bea (l_{(f_0,p_0)})_*\circ\tau\circ (r_{(f_0^{-1},1/(p_0(f_0^{-1})))})_*(X,p')&=&
 (l_{(f_0,p_0)})_*\circ(\frac{d}{dt}_{|t=0}\xi_g(f_t,\frac{p't+p_0}{p_0}(f_0^{-1}))\nn\\
 &=& f_0^*(p'g+p_0L_Xg)\eea
 This is equivalent to the following:
\be\label{equivariant transf}T_{(f_0,p_0)}\xi_g(X\circ f_0,(p'\circ f_0)p_0)=p_0f_0^*\tau_g(X,p') \ee

 Hence we find that for the conformal group every structure works nicely and we have a metric-induced connection
 in $\mathcal{C}\hookrightarrow\M''\ra \M''/\mathcal{C} $ for every choice of $\beta$ in the supermetric
and positive lapse potential. The actual equations, writing $\omega[\dot g]=((\omega_{\DD}[\dot g])^a,
  \omega_{\mathcal{P}}[\dot g])$
  take the  form
\bea\label{conformal connection 1} (\omega_{\DD}[\dot g])^a_{~;a}+3\omega_{\mathcal{P}}[\dot g])=\dot g^a_{~a}\\
\label{conformal connection 2}\left(N((\omega_{\DD}[\dot g])^{(a;b)}+\omega_{\mathcal{P}}[\dot g]g^{ab}-\dot
g^{ab})\right)_{;b}=0
 \eea
meaning the corrected velocities are both traceless and transverse {\it with respect to the positive definite
(ultra-local) metric in $\M$} given by $G^{abcd}=Ng^{ac}g^{bd}$.

Thus we have guaranteed existence and uniqueness of solutions for the connection form for $\mathcal{C}$.

\subsection{A fully conformally invariant action with two metric degrees of freedom.}

\subsubsection{Horava gravity with detailed balance.}

Let us briefly examine one recent gravity theory which breaks foliation invariance and possesses powerful indications towards conformal invariance. The so-called Horava-Lifshitz gravity has recently received a great deal of attention, we present its ``detailed balance" formulation \cite{Horava:lif_point}:
\be\label{equ:Horava_action}S=\int dt\int d^3 x\sqrt g N\left(\frac{2}{\kappa^2}G_\lambda^{abcd}K_{ab}K_{cd}-\frac{\kappa^2}{2w^4}C^{ab}C_{ab}\right)
\ee
where $w$ and $\kappa$ are coupling constants. The cotton tensor used here is defined as
\be\label{equ:cotton_tensor} C^{ab}:=\epsilon^{acd}\nabla_c\left({R^b}_d-\frac{1}{4}\delta^b_d R\right)
\ee
The Cotton tensor is symmetric, transverse and traceless:
\be\label{equ:cotton_tensor_props} C^{ab}=C^{(ab)}~~,~~ {C^{ab}}_{;b}=0~~,~~{C^a}_a=0
\ee
It also homogeneously scales conformally with weight $-5/2$. Thus under $g_{ab}\rightarrow e^{4\phi}g_{ab}$ we get $C^{ab}\rightarrow e^{-10\phi}C^{ab}$.

  We shall not explain the more interesting aspects of \emph{why} the action \eqref{equ:Horava_action} was introduced in the first place. At this level, it suffices to say that it possesses different numbers of spatial and time derivatives, making it space-time anisotropic and power-counting renormalizable.
   By the previous work on
  this section we need not check that the constraints associated to the action of $\DD$ and $\mathcal{C}$ (the ``diffeomorphism" and ``conformal" constraints) propagate, have the right transformation properties, or any
  of a multitude of laborious computations; if the system is consistent, we have designed a well-defined gauge system, in every possible sense.
  Thus all that is required for conformal invariance is to impose equations \eqref{conformal connection 1}
  and \eqref{conformal connection 2} for our fixed connection form.
  The remaining constraint, the scalar constraint of theory \eqref{equ:Horava_action} is of the form:
  \be\label{Hamiltonian constraint1} \frac{2}{\kappa^2}K_{TT}^{ab}K^{TT}_{ab}-\frac{\kappa^2}{w^4}C^{ab}C_{ab}=0
  \ee
 where $NK_{TT}=\dot g_{TT}$ and
 \be\label{corrected}\dot g^{TT}_{ab}=(\dot g_{ab}+(\omega_{\DD}[\dot g])_{(a;b)}+\omega_{\mathcal{P}}[\dot
g]g_{ab})\ee are the traverse traceless metric velocities.

Let us proceed to count the degrees of freedom of the  conformally invariant system we expect to obtain (if the action is consistent).
  First of all, since we no longer have full Lorentz invariance, the modified
  version of the Hamiltonian constraint \eqref{Hamiltonian constraint1},
  does not automatically propagate, yielding one further constraint on our
   variables. Thus we have: $6+6+1=13$ (degrees of freedom in $g_{ab}, \dot g_{ab}$ and $N$) minus `the equations of
    motion for N (or `Hamiltonian constraint') and its propagation equation,
    which are (at least) 2 in number, the `conformal constraints on corrected
    velocities' \eqref{conformal connection 1} and
  \eqref{conformal connection 2} (which gives $4$ more), and finally the additional $4$ coming from the choice of
  section for $\mathcal{C}$. Thus we have $3$ remaining degrees of freedom, which is one too little.

{\flushleft{ \bf A model with the right number of degrees of freedom.}}\smallskip

First of all, we see that to make a fully $\mathcal{C}$-invariant theory, we must have a local lapse, and if so,
to get the right number of degrees of freedom the lapse equations of motion have to be automatically propagated by
the other constraints and equations of motion, which is a very tall order.

An alternative, which we propose here, does not have a lapse, and thus has no Hamiltonian constraint
 and yields a theory with the right number of degrees of freedom (12-4-4=4). For this we simply
choose the following lapse potential, satisfying \eqref{lapse potential 2}:
\be\label{Cotton}N_g(x)=\sqrt{C^{ab}(x)C_{ab}(x)}\ee Its conformal weight is given by $\frac{-20+8}{2}=-6 $, and it clearly satisfied all the items of  Definition \ref{lapse potential}) and thus all our gauge constructions are valid. We thus have the action: \be\label{new action} S= \int dt\int_M
\sqrt{g}d^3x \frac{1}{w^2}\sqrt{C^{ab}C_{ab}}(\dot g^{cd}_{TT}\dot g^{TT}_{cd})
  \ee

To stress, this is a fully $\mathcal{C}$-invariant action with the same number of degrees of freedom as GR, but
which does {\it not} have a Hamiltonian constraint.  Equation  \eqref{new action} is furthermore a purely geodesic-type action in Riem, with just one global lapse and thus one global notion of time, as such it also possesses inherent value in a relationalist setting.

In the first part of this thesis we were able to see GR fully as conformally invariant theory. So if any, this formalism has hopes only of
recovering this dual formulation of GR, something which will be investigated further.

\section{Discussion}\label{Conclusion}

We have  started a study of  the link between Machian relationalism and gauge theory. To do so, we exploited
the formal gauge structure that the space of  metrics  without symmetries $\M'\subset\M$ has. That is, under the
action of the diffeomorphism group $\DD$, $\M'$ is a principal fiber bundle, and so can  be made to harbor the
usual related constructions: connection, curvature, etc. In spite of some natural deviations, we found that the
majority of the structures present in gauge theory can be suitably transplanted to this infinite-dimensional
setting. This is due mostly to the existence of a cross-section of $\M'$ relative to $\DD$.  This gauge structure
naturally embodies the freedom to `label' space, and hence should be fully treated as the natural gauge structure
that reflects relationalism.

The connection should satisfy the usual properties required in ordinary
finite-dimensional gauge theory. Utilizing some
background from the theory of the spaces of maps, we have provided a valid working definition of a connection
induced by supermetrics on $\M'$. We have shown rigorously under which conditions  the combined action of the
gauge group we are utilizing and choice of supermetric on $\M'$ formally define a connection.

The connection turns out to be a vector-field valued distribution, locally represented as:
$\omega\in\Gamma(TM)\otimes\Gamma(TM\otimes_S TM)$. Basically, the local form will define the way a given
metric velocity at a given point $y$, will contribute to the best-matching field at another point, $x$.

To emphasize, the use of the connection form is to provide an identification of (spatial) points of $M$ along time
(or more appropriately, along curves in $\M'$). For the conformal theory, in conjunction to the 3-diffeomorphisms, it would also mean that one finds a way to provide an identification of scale along time. In this way, it is naturally related to Barbour's best-matching
procedure \cite{Barbour94} and Wheeler's geometrodynamical setting of the 3+1
decomposition \cite{BSW}.

 The more interesting prospects for the theory are now to further investigate the structure of
 connection forms for groups such as that of conformal transformations, for which the connection seems
 much better behaved than that of the $\diff$ group.

The non-projectable Horava action  is
      tantalizingly close to being fully $\mathcal{C}$-invariant. All we require to make it completely so  is
      to impose the full connection of $\mathcal{C}$ on its velocities \eqref{conformal connection 1},
      \eqref{conformal connection 2}. Having done so however takes us to a theory with only 3 degrees of
      freedom. There exists a modification of Horava which allows us to recover the right count \eqref{new
action}. Applying our results on the gauge structure of $\mathcal{C}$, we have a completely well defined theory
given by the action \eqref{new action}, all constraints automatically propagate by the equivariance properties of
the connection, for which existence and uniqueness is guaranteed. But of course, having the right number of degrees of
freedom does not guarantee us to recover GR in the right limit.

 Another
possible avenue of research is to study quantum gravity from the perspective of a gauge field theory over $\riem$,
for which many of the structures presented here are not sufficient and will need to be expanded.

We leave the following inter-related questions for further study:
 \begin{itemize}
 \item {\bf A Kaluza-Klein interpretation of the  forcing terms that appear in the 3+1 dynamical equations.}
 - I.e. the dynamical Einstein equations cannot be given exactly by some geodesic equation for some metric
 which projects down to $\super$, there are so called forcing terms that deviate. Can these terms can be seen
 as originating from a specific connection (or more specifically, curvature) in the PFB given by $\M$?
     \item
       {\bf Further study of spontaneous symmetry breaking from the full conformal invariant theory  to the
       merely $\diff$ invariant one.} - To be more specific, under the action of the conformal group, $\M$
       also has a natural dense subset over which it can be given a PFB structure. This action in fact has
       much nicer properties and so far none of the pathologies contained in the $\DD$ case. Furthermore, from
       interesting splitting theorems present for the conformal group  the
       orbits of the conformal group decomposes into irreducible elements given by  different sets of
       subgroups. This is an interesting setting then to study spontaneous symmetry breaking from the
       conformal group to the merely $\diff$ invariant one, and enquire over the properties of some unified
       theory possessing the full conformal group as gauge. This also ties in with Horava gravity and with
       field theory as seen below.
       \item {\bf The specific relationship with the dual volume-preserving conformal theory}-What is the relationship between the dual theory presented in \cite{Gomes2011a} and the conformal action given in \eqref{new action}? What happens if one tries to do gauge theory for the group of volume-preserving conformal transformations, as is done in \cite{Gomes2011a}? Does one go back to that action? What are the uniqueness results?
      \item {\bf The specific relationship with Horava gravity.}-
       If the modification of Horava which allows us to recover the right count \eqref{new action}, does not
yield GR in the right limit, using the splitting properties of $\mathcal{C}$, is there a subgroup which may do
the trick?
              \item {\bf Field theory over $\M'$. }
        - The connection, or potential, in electromagnetism, cannot be dynamically determined for actions over
        particle trajectories. To dynamically determine the connection one needs curvature terms included in
        the action, and of course the pull-back of a two-form over a curve is bound to be zero. In the same
        way, each history of the Universe is here a trajectory in $\M$ (or $\super$ if you will), and any
        curvature term of $\M'$ as a PFB will automatically vanish. Hence either we take the connection as
        given, as we have done in this paper for the specific case of the supermetric given connection, or we
        have to expand our theory to a field theory over $\M$, and therein we can define a non-trivial
        curvature term and an appropriate action containing it. Going back to the first item on this list one
        could now of course perform the full Kaluza-Klein analysis and also vary for the possible
        supermetrics, finding possibly  $\beta\neq1$\footnote{By symmetry requirements the supermetric
        variation will amount to varying with respect to the parameter $\beta$ in the DeWitt metric}. This is
        also exactly analogous to what happens in Horava's theory and thus ties in with all of the above
        items.
         \end{itemize}

\paragraph{Acknowledgments}I would like to thank Julian Barbour for stimulating me to pursue this work and providing an inspiration. I would also like to thank besides himself, Sean Gryb, Tim Koslowski and Ed Anderson for countless discussions on the subject.

\appendix
 \begin{center}
    {\bf APPENDIX}
  \end{center}
The important result for a gauge theory in $\M'$  is the Ebin-Palais  slice theorem \cite{Ebin}. It is analogous
to the usual slice theorem, and it is that which reveals the principal fiber bundle structure in $\super'$. We describe necessary material for the construction of a principal connection in $\M'$, with the principal aim
being achieved in  Theorem \ref{proj theo}. But to see why the analogy between the free action of the $\DD$ group on
$\M'$ and finite-dimensional principal fiber bundles is more than an analogy we refer the reader to \cite{Michorbook}.

The material in the first section follows to a certain degree \cite{Ebin}, but for the reader's convenience we
give a description in our language of the material necessary for us. I.e. the material necessary for rigorously
defining and constructing the connection through the use of a metric in $\M$.
Furthermore, the equivalent conditions on Theorem \ref{proj theo} are new and are one of the main characters in
the main text.

\section{Gauge structures over Riem: Slice theorem.}\label{sec:slice}

The important result for a gauge theory in $\M'$  is the Ebin-Palais  slice theorem \cite{Ebin}. It is analogous
to the usual slice theorem, and it is that which reveals the principal fiber bundle structure in $\super'$. We describe necessary material for the construction of a principal connection in $\M'$, with the main aim
being achieved in  Theorem \ref{proj theo}. But to see why the analogy between the free action of the $\DD$ group on
$\M'$ and finite-dimensional principal fiber bundles is more than an analogy we refer the reader to \cite{Michorbook}.

To a certain degree the material in the first section follows  \cite{Ebin}, but for the reader's convenience we
give a description in our language of the material that we need, i.e., the material necessary for the rigorous
definition  and construction of the connection through the use of a metric in $\M$.

\subsection{Constructing the vertical projection operator for the PFB-structure of $\M'$}\label{sec:constructingV}
The constructions here include technicalities needed in order to define the spaces we work with as proper Hilbert
manifolds, in order that we can use certain theorems only applicable in that domain.  We use these Hilbert spaces and the
Riemannian metric in the third subsection, to define the structure of the $\DD$ orbits in $\M$. It is here that we
define and use the Fredholm alternative most intensely. The bundle normal
to the orbits (called the horizontal bundle in the main text) and the orthogonal projection with respect to such
a decomposition is constructed.
Hence this is the section used in the following chapter in the construction of the principal connection $\omega$ on $\M'$,
based on the existence of a metric on the Hilbert completion $\M^s$ (see below). Although the constructions here
are based in $\M^s$, it can be shown that they can be later transported to our merely $C^\infty$ setting of $\M'$
\cite{Ebin}. Lastly we state and sketch the remaining steps in the proof of the slice theorem, on which the whole
gauge apparatus is based.

\subsubsection{$H^s$-manifolds. Sobolev Lemma and all that}

Suppose that $E$ is a vector bundle over a smooth closed manifold $M$;  $\pi_{\mbox{\tiny E}}:E\ra M$. \begin{itemize}
\item Let $\Gamma^k(E)$ be the space of $k$-differentiable sections of $E$, this is a Banach space with topology of uniform
convergence up to $k$ derivatives.
\item Let $J^s(E)$ be the $s$-th jet bundle of $E$, which we endow with (for now) any Riemannian structure
$\langle\cdot,\cdot\rangle_s$. For a fixed volume element of $M$, let us call it $d^3x$, we get the inner product
on the space of sections  $\Gamma^\infty(J^s(E))$ by $$(a,b)_s=\int_M\langle a,b\rangle_s d^3x$$ Since there is a
natural linear map from $\Gamma^\infty(E)$ to $\Gamma^\infty(J^s(E))$ (basically given by successive
linearizations), this also defines an inner product on $\Gamma^\infty(E)$. Now we define
$$ H^s(E)\mbox{~~~is the completion of~~}~\Gamma^\infty(E) \mbox{~~~with respect to~~~} (\cdot,\cdot)_s
$$ As such it is a Hilbert space whose norm depends on the choices of inner product and volume form,
but whose topology does not. In local coordinates, this is the space of sections of $E$ which in local coordinates
have partial derivatives up to order $s$ square integrable, i.e. for $f\in H^s(E)$ the norm is given in local
coordinates by
$$ ||f||_{s}=\sum_{0\leq\alpha\leq s}||\partial^\alpha
f||_{L^2}=\sum_{0\leq\alpha\leq s}\sqrt{\int_M|\partial^\alpha f |^2d^3x}
$$ We note in passing that for $p\neq2$ the above is not a Hilbert space for the $L_p$ norm.
 \end{itemize}

Now to construct the appropriate manifolds, we will need the following
\begin{lem}[Sobolev Lemma]For $n=\mbox{dim}(M)$, if $s>k+n/2$ we have that $H^s(E)\subset\Gamma^k(E)$ and the
inclusion is a {\it linear continuous} map.
\end{lem} Note that the lemma is very far from trivial, since, of course we always have
$\Gamma^{k+1}(E)\subset\Gamma^k(E)$, but the $s$-th completion of the $\Gamma^\infty(E)$ sections could have
elements that were not smooth.

\subsection{Defining $\M^s$, a Riemmanian structure for $\M^s$, and an $\exp$ map.}

Let $E=S^2T^*:=T^*M\otimes_ST^*M$, the symmetric product of the cotangent bundle. The space of positive definite
smooth sections of $S^2T^*$ is what we call $\M$. i.e. $\M=\Gamma^\infty_+(S^2T^*)$. Abusing notation, let
$\Gamma_0(\M):=\Gamma_+^0(S^2T^*)\subset\Gamma^0(S^2T^*)$ be the space of merely continuous metrics on $M$, which
is an open subset of $\Gamma^0(S^2T^*)$. The set $\Gamma_0(\M)$ still is only endowed with a topology. To make it
into the appropriate Hilbert manifold, we define
$$\M^s:= H^s(S^2T^*)\cap\Gamma_0(\M)
$$ Now, by the Sobolev lemma,  the inclusion $\iota:H^s(S^2T^*)\hookrightarrow \Gamma^0(S^2T^*)$ is continuous for $s>1$ in $n=3$.
Since $\Gamma_0(\M)$ is an open subset of $\Gamma^0(S^2T^*)$, we have that $\M^s=\iota^{-1}(\Gamma^0(\M))$ is an
open set in $H^s(S^2T^*)$ and hence a Hilbert manifold. A similar construction is available to transform the group
of diffeomorphisms $\DD$ into a Hilbert manifold $\DD^s$, but as we will not get into the intricacies of the last
part of the proof of the Ebin-Palais section theorem, we will not need it, and hence just use the generic
$\Gamma(TM)$ as the tangent space to the identity of $\DD$.

For each point of the Hilbert manifold $\gamma\in\M^s$ we have that $\gamma$, being an inner product on $TM$,
induces an inner product in all product bundles over $TM$, and hence we have an induced inner product on $S^2T^*$,
which we call $\langle\cdot,\cdot\rangle_\gamma$. It furthermore induces a volume form, and thus we have the
induced inner product on each $T_\gamma\M^s\simeq H^s(S^2T^*)\ni \alpha, \beta$. \be\label{section
ip}(\alpha,\beta)_\gamma=\int_M \langle\alpha,\beta\rangle_\gamma d\mu_\gamma \ee Since $\M^s\subset\Gamma^0(\M)$,
$(\cdot,\cdot)_\gamma$ induces the $H^0$ topology on $H^s(S^2T^*)$,  there might be sequences in
$H^s(S^2T^*)$ that converge with respect to $(\cdot,\cdot)_\gamma$ but not to an element $H^s(S^2T^*)$. This is what we mean when we say that
$(\cdot,\cdot)_\gamma$ is  merely a weak Riemannian metric on $\M^s$.\footnote{ This sort of lack of metric convergence in $H^s$, poses certain issues when objects are only implicitly defined by the metric.}

For $f\in\DD$, as extensively used in the main text, $f^*:\M^s\rightarrow \M^s$ acts linearly, so furthermore
$T_\gamma f^*=(f^*)_{*}=f^*:H^s(S^2T^*)\rightarrow H^s(S^2T^*)$. From the properties $\langle
Tf^*\alpha,Tf^*\beta\rangle_{f^*\gamma}=\langle\alpha,\beta\rangle_\gamma\circ f$ and $d\mu_{f^*g}=f^*d\mu_g$ it
is straightforward to show that $(\cdot,\cdot)_\gamma$ is $\DD$-invariant.

\subsection{The orbit manifold and splittings}\label{orbit manifold section}

Consider now the map
\begin{eqnarray*}
\Psi:\M^s\times \DD &\ra& \M^s \\
~(g,f)&\mapsto & f^*g
\end{eqnarray*}
As in the previous section,  the image of $\Psi_g$, $\mathcal{O}_g=\Psi_g(\DD)$ is called {\it the orbit of $\DD$ through $g$}.
We have that the derivative of the orbit map $\Psi_g: \DD \ra \M$ at the identity, which we will call $\alpha_g$:
 \be\label{equ:def:alpha}\alpha_g:=T_{\Id}\Psi_g: X\mapsto
L_Xg=\imath_X(L_\cdot g)\ee where $X\in \Gamma(TM)$ is the infinitesimal generator of a given curve of
diffeomorphisms of $M$. We may also write $\alpha_g$ as $\alpha_g(X)=(T_{(g,\Id)}\Psi)\cdot(0,X)$, which may make the meaning of the map more clear. It takes each element of the Lie algebra into its fundamental vector field, i.e. it gives directions along the orbits corresponding to certain directions along the group.

 We want to calculate what $T_{f}\Psi$ is with respect to
$T_{\Id}\Psi$. For $\eta, f\in \DD$ and $r_f$  the right action of diffeomorphisms (for which $T
(r_f)=(r_f)_{*}:\Gamma(TM)\rightarrow \Gamma(TM)$), we have, $$ f^*\circ\Psi(g,r_{f^{-1}}(\eta))=f^*\circ\Psi(g,\eta\circ
f^{-1})=f^*(\eta\circ f^{-1})^*(g)=\Psi(g,\eta) $$ therefore \be\Psi=f^*\circ\Psi\circ r_{f^{-1}}\ee and thus
\be\label{orbit transf} T_f\Psi=T_gf^*\circ T_{\Id}\Psi\circ (r_{f^{-1}})_*=f^*\circ\alpha\circ
(r_{f^{-1}})_*:T_f\DD\rightarrow H^s(S^2T^*)\ee This equation is of course equivalent to saying that at $f$
$$T_{f}\Psi_g(X\circ f)=f^*\alpha_g(X)
$$

Since the maps above are isomorphisms, we conclude from \eqref{orbit transf} that $ T_f\Psi(T_f\DD)$ is isomorphic to $
T_{\Id}\Psi(T_{\Id}\DD)= T_{\Id}\Psi(\Gamma(TM))$, and thus all tangent spaces to the orbits are isomorphic.

For a  finite dimensional vector space $E$, we can always algebraically split a subspace $F_1$ from its complement $F_2=F_1^C$. For infinite-dimensional vector spaces, a closed finite-dimensional subspace also always has a closed complement subspace. In the general case of closed infinite-dimensional subspaces though, the complement $F_1^C$ of $F_1$ is not necessarily  closed, and upon closure it might not be in the complement.

Now we have to show that the tangent space to the orbits splits. I.e. that not only is the image of
$T_{\Id}\Psi=\alpha$ a closed linear subspace of $H^s(S^2T^*)$, but also that it has a closed complement $\mbox{Im}\alpha)^{\mbox{\tiny C}}$ and thus
 $H^s(S^2T^*)\simeq\mbox{Im}\alpha\oplus(\mbox{Im}\alpha)^{\mbox{\tiny C}}$. We will do this in the following
detour through functional analysis.

\subsubsection{Splitting of $T\M$ by $T\mathcal{O}$.}

In local charts of $E$ and $F$,  for $E$ and $F$ vector bundles over $M$, a $k$-th order differential operator
$D:\Gamma^\infty(E)\ra \Gamma^\infty(F)$, acting on $f\in \Gamma^\infty(E)$  can be written as \footnote{Here we
use $f$ to make the analogy with vector-valued functions in local charts more transparent}:
$$ D(f)=\sum_{0\leq|i|<ka_i}a^i\frac{\partial^{|i|}f}{\partial x^{i_{\mbox{\tiny 1}}}\cdots \partial x^{i_{\mbox{\tiny
n}}}}
$$ where $i=(i_{\mbox{\tiny 1}},\cdots,i_{\mbox{\tiny n}})$, $n=$dim$M$ and $|i|=\sum i_{\mbox{\tiny n}}$
and $a^i(x)\in L(E_x,F_x)$.

For each $x\in M$ and for ${\bf p}\in T_x^*M$, the symbol of an operator $D$ is a linear map $\sigma_p(D):E_x\ra
F^*_x$. Basically what one does, in local coordinates, is to replace the highest order partial derivatives by the
components of $p$: $\partial/\partial x^i\ra p_i$.  The symbol of a differential operator will be said to be
injective if the resulting linear operator is injective.

The $k$-th order differential operator $D:\Gamma^\infty(E)\ra \Gamma^\infty(F)$ trivially extends uniquely to a
continuous linear map between the Hilbert spaces $D:H^s(E)\ra H^{s-k}(F)$. If inner products
$\langle\cdot,\cdot\rangle_E, \langle\cdot,\cdot\rangle_F$ in $E$ and $F$ respectively, together with a measure
for $M$ are given, we call $(\cdot,\cdot)_E, (\cdot,\cdot)_F$ the inner products induced in $H^s(E)$ and
$H^{s-k}(F)$ respectively. By the Riesz representation theorem, there then exists a unique adjoint for any such
$D$: \be\label{adjoint D}(a,Db)_E= (D^*a,b)_F \mbox{~~~for~~~}a\in H^s(E), b\in H^{s-k}(F)\ee

Now, a well-known theorem in functional analysis tells us that, if a differential operator is elliptic  it
possesses the splitting property :
\begin{theo}[Fredholm Alternative]\label{theo:Fredholm}Let $D$ be an elliptic differential operator of $k$-th order,\footnote{
There are subtleties here regarding the order $s$ of the Sobolev spaces in each side \cite{Ebin}, but these do not
concern us here. For the avid reader, the order of the spaces can be worked out by the Regularity theorem , which
states that for an elliptic operator of order $k$, and $f\in L_2(E)$, $D(f)\in H^{s-k}$ implies $f\in H^s$. The
Weyl lemma, stating that if the Laplacian (which is an elliptic operator) of an $L_2$ function is zero (and zero
is in $H^s$ for any $s$) then the function $f$ is $C^{\infty}$, is an immediate corollary of the regularity
theorem.} then
 \be\label{Splitting}
H^{s-k}(F)=\mbox{Im}(D)\oplus \mbox{Ker}D^*\ee \end{theo}This stems from the more general fact, that for any linear densely-defined (i.e., having a domain of definition that is dense in H)
operator $A$, not necessarily bounded, we have the splitting property:
\be\label{equ:weaker_splitting}A=\overline{\mbox{Im}(A)}\oplus \mbox{Ker}(A^*)
\ee where the overline denotes closure.
  We will not dwell on the proof, we merely mention that
the necessary ingredients are norm bounds in the presence of elliptic operators to show that $\mbox{Im}(D)$ is
closed, and that $\mbox{Im}(D)\bot_{L_2} \mbox{Ker}D^*$ implies an $H^s$ splitting. For a different take on the Fredholm alternative, see \cite{Ramm}.

 The operator $\alpha_g:\Gamma(TM)\ra H^s(S^2T^*): X\mapsto X_{(i;j)}$
 can easily be shown to have injective symbol, since for $p\in T_x^*M$, $v\in T_xM$
 such that $\xi=g(v,\cdot)$ we have $$ \sigma_p(\alpha)(v)=\xi\otimes_{\mbox{\tiny S}} p
 $$ where again the subscript $S$ stands for the symmetrized tensor product.
  Furthermore, since
$\sigma(D^*\circ D)=\sigma(D)^*\circ\sigma(D)$, it follows that if $\sigma(D)$ is injective, then for
positive definite inner product we automatically have $\sigma(D^*)$ surjective and
$\Ker(\sigma(D^*))\cap\mbox{Im}(\sigma(D))=0$ (trivial, see proof of Proposition \ref{propo}).
Then $\sigma(D^*\circ
D)$ is an isomorphism, which by definition  makes $D^*\circ D$, or  in our case, $\alpha^*\circ\alpha$
an elliptic operator. Applying the above equation
\eqref{Splitting} to $\alpha^*\circ\alpha$ we  arrive at
$$\Gamma(TM)=\mbox{Im}(\alpha^*\circ\alpha)\oplus \mbox{Ker}(\alpha^*\circ\alpha)$$
from which we conclude that $\alpha^*\circ\alpha: \mbox{Im}(\alpha^*\circ\alpha)\ra \mbox{Im}(\alpha^*\circ\alpha)$ is
an isomorphism.

We will now sketch how under the present conditions, using ellipticity of $\alpha^*\circ\alpha$, a similar splitting
automatically applies
for $D=\alpha$.
\begin{prop}\label{propo}If $D^*\circ D$ elliptic and  the restricted inner products $(\cdot,\cdot)_{E_{| \mbox{\tiny Im}(D)}}$ and $(\cdot,\cdot)_{F_{| \mbox{\tiny Im}(D^*)}}$ are non-degenerate, then  $\mbox{Ker}(D^*\circ D)=\mbox{Ker}D$ and $\mbox{Im}(D^*\circ
D)=\mbox{Im}D^*$ which implies
 \be H^s(S^2T^*)=\mbox{Im}(D)\oplus
\mbox{Ker}(D^*)\ee for $D=\alpha$.\end{prop}
 {\rm Proof.}
That $\mbox{Ker}(D^*\circ D)\supset \mbox{Ker}D$ is clear. Now suppose $a\in H^s(E), c\in H^{s-k}(F)$, then if
$D^*\circ Da=0$, we have $(Da,Db)_F=0$ for all $b\in  H^s(E)$ which implies $Da=0$ if the inner product restricted to Im$(D)$ is non-degenerate. This shows $\mbox{Im}(D)\cap
\mbox{Ker}D^*=0$.

Also $\mbox{Im}(D^*\circ D)\subset\mbox{Im}D^*$ from the
outset. To show $\mbox{Im}(D^*\circ D)\supset\mbox{Im}D^*$,  since $H^{s-k}(E)=\mbox{Im}(D^*\circ D)\oplus
\mbox{Ker}(D^*\circ D)$ and $\mbox{Ker}(D^*\circ D)=\mbox{Ker}D$ we have merely to show that  $\mbox{Im}(D^*)\cap
\mbox{Ker}D=0$. Suppose $b\in \mbox{Im}(D^*)\cap \mbox{Ker}D$, i.e. $b=D^*c$ and $Db=0$, then  $(D^*c,D^*d)_E=0$ for all $d\in H^{s-k}(F)$.
Then if the inner product in $E$ restricted to $\mbox{Im}(D^*)$ is non-degenerate,  $D^*a=b=0$. Thus we have proved the first part of the proposition.

Now for the second part we already have  $\mbox{Im}(D)\cap \mbox{Ker}D^*=0$; to show that $H^s(S^2T^*)$
is generated by $\mbox{Im}(\alpha)+\mbox{Ker}(\alpha^*)$ we write:\footnote{Even though we have only shown the above
direct sum exists in the linear algebraic sense, the closed graph theorem guarantees it extends to
the topological domain.},
$$H^s(S^2T^*)=(D^*)^{-1}(\mbox{Im}(D^*))=(D^*)^{-1}(\mbox{Im}(D^*\circ D))=(D^*)^{-1}(D^*\circ D(H^s(TM))$$ Since
$\mbox{Im}(D)\cap \mbox{Ker}D^*=0$  $$(D^*)^{-1}(D^*\circ D(H^s(TM)))=H^s(S^2T^*)=\mbox{Im}(D)\oplus
\mbox{Ker}(D^*)~~~~~~~~\square$$

Note  that, in the first part of the proposition,  $\mbox{Im}(D^*\circ D)\supset\mbox{Im}D^*$ is equivalent to
$\mbox{Ker}(D^*\circ D)=\mbox{Ker}D$, if the inner product in $H^s(E)$ is positive-definite. And of course, if $D$
is injective, this is equivalent to $\Ker(D^*)\cap \mbox{Im}(D)=0$, which is the usual equation to define the
orthogonality relation (but not a projection).

Now it is relatively straightforward to show that \eqref{Splitting} is valid for $D=\alpha$, which shows that for
$\M'$, the map orbits are injective immersions. To show that they are also embeddings requires more work, which
again we will not go through since it does not contribute anything to our constructions. Thus omitting the prof we shall, for $g\in{\M'}^s$, take $\Psi_g:\mathcal{O}_g\ra{\M'}^s$ to be an embedding.

\subsection{The normal bundle to the orbits and construction of the vertical projection operator.}\label{projection section}
The  bundle orthogonal to $\mathcal{O}_g$ is defined as \be\label{ortho bundle}\nu(\mathcal{O}_g):=\{n\in
T\M^s_{|\mathcal{O}_g}~|~(n,v)=0, \mbox{~~for~~}v\in T\mathcal{O}_g\}\ee Given a Riemannian structure on $\M^s$,
the bundle orthogonal  with respect to it would  automatically be a smooth subbundle, however we possess so far
merely a weak Riemannian metric, and so must put in a little more effort.

From the previous subsection we have seen that for any $g\in \M'$,  there exists an isomorphism
\be\label{Splitting alpha}T_g\M\simeq H^s(S^2T^*)\simeq\mbox{Im}\alpha\oplus \mbox{Ker}(\alpha^*)\ee Hence, since
for $v\in\mbox{Ker}(\alpha^*)$ it follows that $(\alpha(X),v)=0$ and $\mbox{Im}(\alpha_g)\simeq T_g(\mathcal{O}_g)$
we have that $\nu(\mathcal{O}_g)_h=\mbox{Ker}(\alpha_h^*)$.

 We shall thus define a smooth, surjective
map: $P: T\M^s_{|\mathcal{O}_g}\ra T\mathcal{O}_g$, such that $\Ker(P)=\mbox{Ker}(\alpha^*)=\nu(\mathcal{O})$
which will turn out to be exactly the vertical projection $\hat V$ we need for the definition of the principal
connection $\omega$. Before proceeding we note that in finite dimensions an orthogonal projection operator can be
easily defined from a basis, but that in the present case an orthogonality relation  does not automatically define a projection, even for a positive definite inner product.

From Proposition \ref{propo} we have  Im$(\alpha^*\circ\alpha)=\mbox{Im}(\alpha^*)$; hence for each point
$g\in\M$, $\alpha^*(H^s(S^2T^*))=\alpha^*\circ\alpha(\Gamma(TM))$. From the above consideration we can regard
$\alpha^*\circ\alpha_{|\mbox{Im}(\alpha^*)}$ as a map from $\mbox{Im}(\alpha^*\alpha)$ to itself, which,  from
self-adjointness and ellipticity, means it is in fact an isomorphism, having thus a smooth inverse.
 Hence we define: \be\label{P vertical}P:=\alpha\circ(\alpha^*\circ\alpha)^{-1}\circ\alpha^*:H^s(S^2T^*)\ra H^s(S^2T^*)\ee
It is clear that  $P^2=P$,  that $\nu(\mathcal{O}_g)_h=\mbox{Ker}(\alpha_h^*)=\mbox{Ker}P_h$, and that for a
vertical vector, i.e. $v=\alpha(X)$, we get
$P(v)=\alpha\circ(\alpha^*\circ\alpha)^{-1}\circ\alpha^*\alpha(X)=\alpha(X)$, hence the projection acts as the identity on the
vertical space. Thus the following decomposition holds\footnote{ Again, to go from merely
algebraic decomposition to topological decomposition, one must use the closed graph theorem, which says that for
Banach spaces $A, C$ then for a continuous linear operator $f$ such that $f(A)$ is a closed subspace, then  there is a
closed complement $B$, such that $C=f(A)\oplus B$.}: $W=\mbox{Im}T\oplus\Ker T $.
 and thus:
 $$
 H^s(F)=\mbox{Ker}(P)\oplus \mbox{Im}(P)$$
  All that is left to do is check the transformation properties of
$P$.

Let us recall first of all that $\alpha=T_{\Id}\Psi$, and from \eqref{orbit transf}
$$ \alpha_f=f^*\circ\alpha\circ ((r_{f})_*)^{-1}\mbox{~~~and~~~}\alpha^*_f=(r_{f})_*\circ\alpha^*\circ(f^*)^{-1}
$$ thus we can prove the \emph{equivariance} of $P$:
\bea \alpha_f\circ(\alpha_f^*\circ\alpha_f)\circ \alpha_f^* &=&
 f^*\circ\alpha\circ ((r_{f})_*)^{-1}((r_{f})_*\alpha^*\circ\alpha\circ
  ((r_{f})_*)^{-1})(r_{f})_*\circ\alpha^*\circ(f^*)^{-1}\nn\\
  &=& f^*(\alpha\circ(\alpha^*\circ\alpha)^{-1}\circ\alpha^*)(f^*)^{-1}\label{equivariant V}\eea
Since $\alpha_f$ is automatically smooth, all that is left to check is that $\alpha^*_f$ is smooth, since
$\alpha^*_f\circ\alpha_f$ is an isomorphism and the  inverse map in the restricted Banach space is smooth. We
shall not perform this calculation, which stems directly from the construction of the adjoint. Thus we have proven
the following theorem\footnote{We have actually proven it for the Hilbert extension $\M^s$, but it is shown in
\cite{Ebin} how these constructions can be more or less straightforwardly translated to the $C^\infty$ setting.}
\begin{theo}\label{proj theo}
Given a $\DD$ invariant positive definite metric in $\M'$,   the operator \be
P:=\alpha\circ(\alpha^*\circ\alpha)^{-1}\circ\alpha^*:\Gamma^\infty(S^2T^*)\ra \Gamma^\infty(S^2T^*)\ee has
the following properties: \begin{itemize}\item $\DD$-equivariant. \item $P^2=P$ and $H^s(F)=\mbox{Ker}(P)\oplus \mbox{Im}(P)$
\item $P(\alpha(X))=\alpha(X)$.
\item $\nu(\mathcal{O}_g)_h=\mbox{Ker}(\alpha_h^*)=\mbox{Ker}P_h$. \end{itemize}
\end{theo} So we call it {\it the vertical projection
operator} for this metric.

Let us go through the exact structures that were needed for this theorem (and were implied by positive-defiteness
of the $H^s(E)$ and $H^s(F)$ inner products ).
\begin{itemize} \item The adjoint of the operator $\alpha$  exists and is
smooth (in the main text $\alpha=T_{\Id}\Psi$ where $\Psi:\DD\times\M'\ra\M'$ is the group multiplication
operator).
\item  $\alpha^*\circ\alpha$ is elliptic (which can be checked by its symbol). Then from self-adjointness and the
decomposition \eqref{Splitting} we concluded that $\alpha^*\circ\alpha_{|\mbox{Im}(\alpha^*\circ\alpha)}$ was an
isomorphism. \item Im$(\alpha^*\circ\alpha)=\mbox{Im}(\alpha^*)$, which allowed us to regard
$\alpha^*\circ\alpha_{|\mbox{Im}(\alpha^*)}$ as a map from $\mbox{Im}(\alpha^*\alpha)$ to itself, which meant
$\alpha^*\circ\alpha_{|\mbox{Im}(\alpha^*)}$ was in fact an isomorphism, having thus a smooth inverse. Note that
for this, from Proposition \ref{propo}, we needed only that $\Ker(\alpha^*)\cap\mbox{Im}(\alpha)=0$ and
$\langle\cdot,\cdot\rangle_E$ be positive definite. Thus the injectivity of $\alpha$, combined with the previous item says
$\mbox{Ker}(\alpha_h^*)=\mbox{Ker}P_h$.
\item The metric in $H^s(T^*M\otimes T^*M)$ is $\DD$ invariant. We used to derive the transformation
properties of $P$.\end{itemize}

\subsection{The Slice Theorem}\label{subsec:slice} Since we have come
this far into the constructions in a reasonable degree of detail,
we now state
\begin{theo}[Slice for $\M/\DD$, \cite{Ebin}]\label{theo:slice}
For each $g\in\M$ there exists a contractible submanifold $\Sigma$ of $\M$ containing $g$ such that
\begin{enumerate}
\item $f\in I_gM\Rightarrow f^*\Sigma=\Sigma$
\item $f\notin I_gM\Rightarrow f^*\Sigma\cap \Sigma=\emptyset$
\item There exists a local cross section $\tau:Q\subset \DD/I_g(M)\ra\DD$ where $Q$ is an open neighborhood of the
identity, such that \bea F: Q\times \Sigma &\ra& U_g\\
(f,s)&\mapsto& \tau(f)^*s \eea where $U_g$ is an open neighborhood of $g\in\M$, is a diffeomorphism.
\end{enumerate}\end{theo}
For $\M'$ the space of metrics with no symmetries, the space $\super'=\M'/\DD$  is indeed a manifold
and the existence of a section above allows us to construct
 its local product structure
 $\pi^{-1}(\mathcal{U}_\alpha)\simeq \mathcal{U}_\alpha\times \DD$
through bundle charts and properly {\it define $\M'$ as a PFB}.\footnote{In the  MK sense. See \cite{Michorbook} for an appropriate way to formulate the usual theorems of calculus in this infinite-dimensional setting.}
With this in hand, the usual properties of a principal fiber bundle are proved as in finite dimensions.

\subsubsection*{Remaining gaps in the proof}
For the convenience of the reader we point out the leftover gaps in the proof of the slice theorem. The
steps that we have omitted are $i)$ to take better care of the isotropy group, which we have largely ignored by restricting our attention to the subset of metrics without symmetries (see the notes \cite{gomes-2009} for
a more thorough topological treatment of this subset); $ii)$ the actual construction of a tubular neighborhood for
each fiber using the properties of the exponential map. However, since we have indeed  addressed the major issues
that separate the finite-dimensional case to the present infinite dimensional one, these remaining steps are
closely analogous to the usual finite-dimensional proofs.

 Regarding $i)$, the isotropy group at $g\in \M$ is defined as $I_g:=\{f\in \DD~|~f^*g=g\}$. As $I_g$ is a finite-dimensional, and hence splitting, subspace of $\DD$, all major infinite-dimensional difficulties are more or less
easily dissolved. Since the Lie bracket of vector fields over $M$ commutes with the pull-back by diffeomorphisms,
the distribution of the spaces tangent to $\{I_{f^*g}~|~f\in\DD\}\subset\DD$ is involutive. Hence, using the Frobenius
theorem, we can construct the quotient manifold $\DD/I_g$ and a section for $\pi_{\DD}:\DD\ra \DD/I_g$ on a
neighborhood of the identity, $\chi:U\subset \DD/I_g \ra \DD$. Now define $\Phi_g:\DD/I_g\ra \M$ by
$\Phi_g(I_g\circ f)=f^*g$. Basically we must now replace our results about orbit embeddings for $\Psi$ by the
same results for the effective action, $\Phi$, which is the embedding.

Regarding $ii)$, given a Riemannian metric on a Hilbert manifold, there exists a unique Levi-Civita connection
(which respects both metric compatibility and the no-torsion condition). As we mentioned before, existence of certain objects implicitly defined by a weak metric is not guaranteed, for these objects might lie in the Sobolev completion of the $\mathcal{H}^s$. Thus
uniqueness, but not existence is guaranteed for the Levi-Civita connection. From the two usual coordinate-free
Levi-Civita conditions, using the Jacobi identity one gets for $X,Y,Z$ vector fields on $\M^s$:
$$ (\nabla_XY,Z)_\gamma=\frac{1}{2}\left(X(Y,Z)_\gamma-Z(X,Y)_\gamma+Y(X,Z)_\gamma\right)
$$
We then explicitly calculate the formula above
for three arbitrary vector fields, and upon isolation of the $Z$ vector field on the right hand side find an
explicit expression for the Levi-Civita connection. We will not perform this calculation, which can be checked in \cite{Michor_article}. After this construction we have a smooth exponential map $\exp:T\M^s\ra\M^s$
that is furthermore a local diffeomorphism around the zero section (for fixed base points). Combining this with
the invariance of the metric we get  \bea
f^*(\nabla_XY_{|\gamma})=\nabla_{f^*(X_\gamma)}f^*(Y_\gamma)\\
\exp\circ Tf^*=f^*\circ\exp \eea These relations are instrumental in the building of a section for the action of
$\DD/I_\gamma$.

 We have thus constructed an exponential map for a Hilbert manifold, and we call  the  map $\mbox{Exp}:=\exp_{|\nu(\mathcal{O}_g)}$ the {\it normal
exponential}. It can be seen to be a diffeomorphism
onto a neighborhood of the zero section as follows: the tangent space at a zero normal vector over any point can
be given the  direct sum
 decomposition
 $T_{(g,0)}(\nu\mathcal{O}_g)\simeq T_g\mathcal{O}_g\oplus \nu_g\mathcal{O}_g
$. Over a fixed fiber of $T\M$, i.e. for $v\in{T_g\M}$ , $\mbox{Exp}(g,v)=\exp_g(v)$. We have, taking
$(w,u)=\xi\in T_g\M$,
\begin{eqnarray*}
T_{(g,0)}\mbox{Exp}(\xi)&=&\frac{d}{dt}{|_{t=o}}
\mbox{Exp}(\gamma(t),0)+\frac{d}{dt}{|_{t=o}}\mbox{Exp}(g,tu)=
\frac{d}{dt}{|_{t=o}}(\gamma(t),0)+\frac{d}{dt}{|_{t=o}}\exp_g(tu)\\
~&=&(w,0)+(0,u)=\xi
\end{eqnarray*}
So we have shown that $ T_{(g,0)}(\mbox{Exp})=Id_{|T_g\M}$ which by  the inverse function theorem for Hilbert
manifolds makes the normal exponential a local diffeomorphism that respects the normal decomposition.

Thus  all we now have to do is find a small enough neighborhoods of the zero section of the normal bundle  such
that the $\DD$-transported exponential of some neighborhood of zero on $\nu_h(\mathcal{O}_g)$ satisfies the first
and second item of Theorem \ref{theo:slice}. Finding an appropriate section $\chi:U\subset \DD/I_g \ra \DD$ of the isotropy
group such that the last property is satisfied requires only a small amount of extra work, but it is enough to
make it too much of a detour on the purpose of this section. We refer the reader to \cite{Ebin} for the remaining
details.

It is important to notice that by exponentiating the horizontal subspace at $T_g\M$ one does not necessarily
obtain a horizontal submanifold even for the simple positive definite $\DD$-invariant metric we have used here.
If this were so, there would exist a section in which the connection could be set to zero and thus the curvature of
the connection would automatically vanish. We can see this is not the case since the tool in maintaining
orthogonality through the push-forward of exponential map, the Gauss exponential lemma, only works if one of the
vectors is radial. In other words, in general notation $\langle T_v\exp_pv,T_v\exp_p w\rangle_{\exp_p(v)}=\langle
v,w\rangle_p$ but $\langle T_v\exp_pu,T_v\exp_p w\rangle_{\exp_p(v)}\neq\langle u,w\rangle$. Thus suppose $w$ is
vertical; then it will keep its orthogonality with the radial vector along the exponential, but it will not necessarily
keep orthogonal to the exponential push-forward of another horizontal vector $u\in \HH_g$. It will do this only if
the connection has no curvature.


\end{document}